\documentclass[manuscript]{acmart}
\usepackage{amsmath,amssymb,amsfonts}
\usepackage{algorithmic}
\usepackage{graphicx}
\usepackage{textcomp}
\usepackage{xcolor}
\usepackage{enumitem}
\usepackage{bbm}
\usepackage{multirow}
\usepackage{makecell}
\usepackage{algorithm}

\newcommand{\Uset}{\mathcal{U}}
\newcommand{\Qset}{\mathcal{Q}}
\newcommand{\Cset}{\mathcal{C}}
\newcommand{\Hset}{\mathcal{H}}

\newcommand{\Kset}{\mathcal{K}}

\AtBeginDocument{%
  \providecommand\BibTeX{{%
    \normalfont B\kern-0.5em{\scshape i\kern-0.25em b}\kern-0.8em\TeX}}}

\setcopyright{acmlicensed}
\acmJournal{TOIS}
\acmYear{2025} \acmVolume{1} \acmNumber{1} \acmArticle{1} \acmMonth{1}\acmDOI{10.1145/3716821}

\begin{document}

\title{Rebalancing Discriminative Responses for Knowledge Tracing}

\author{Jiajun Cui}
\email{cuijj96@gmail.com}
\affiliation{%
  \institution{East China Normal University}
  \city{No. 3663, North
Zhongshan Road, Shanghai, 200062}
\country{China}
}
\author{Hong Qian}
\email{hqian@cs.ecnu.edu.cn}
\affiliation{%
 \institution{East China Normal University}
  \city{No. 3663, North
Zhongshan Road, Shanghai, 200062}
\country{China}
}
\author{Chanjin Zheng}
\email{hjzheng@dep.ecnu.edu.cn}
\affiliation{%
 \institution{East China Normal University}
  \city{No. 3663, North
Zhongshan Road, Shanghai, 200062}
\country{China}
}

\author{Lu Wang}
\email{wlu@microsoft.com}
\affiliation{%
 \institution{Microsoft}
\country{China}
}
\author{Mo Yu}
\email{gflfof@gmail.com}
\affiliation{%
  \institution{Tencent}
  \country{China}
}
\author{Wei Zhang}
\authornote{Corresponding author. This work was supported in part by National Key R\&D Program of China (No. 2023YFC3341200), National Natural Science Foundation of China (No. 92270119 and No. 62072182), and Shanghai Institute for AI Education.}
\email{zhangwei.thu2011@gmail.com}
\affiliation{%
 \institution{East China Normal University}
  \city{No. 3663, North
Zhongshan Road, Shanghai, 200062}
\country{China}
}

\renewcommand{\shortauthors}{Jiajun Cui, et al.}

\begin{abstract}
Knowledge tracing (KT) is a crucial task in computer-aided education and intelligent tutoring systems, predicting students' performance on new questions from their responses to prior ones.
An accurate KT model can capture a student's mastery level of different knowledge topics, as reflected in their predicted performance on different questions.
This helps improve the learning efficiency by suggesting appropriate new questions that complement students' knowledge states.
However, current KT models have significant drawbacks that they neglect the imbalanced discrimination of historical responses.
A significant proportion of question responses provide limited information for discerning students' knowledge mastery, such as those that demonstrate uniform performance across different students.
Optimizing the prediction of these cases may increase overall KT accuracy, but also negatively impact the model's ability to trace personalized knowledge states, especially causing a deceptive surge of performance.
Towards this end, we propose a framework to reweight the contribution of different responses based on their discrimination in training. Additionally, we introduce an adaptive predictive score fusion technique to maintain accuracy on less discriminative responses, achieving proper balance between student knowledge mastery and question difficulty. 
Experimental results demonstrate that our framework enhances the performance of three mainstream KT methods on three widely-used datasets.
\end{abstract}

\ccsdesc[500]{Computing methodologies~Neural networks}
\ccsdesc[500]{Applied computing~Education}
\ccsdesc[500]{Information systems~Data mining}

\keywords{knowledge tracing, student behavior modeling, discriminative response rebalance}

\maketitle

\section{introduction}

Due to the rapid development of information technology, online education services in recent decades have led to the availability of extensive teaching materials and student learning information. Consequently, data-driven methods~\cite{data-driven} have gained significant prominence in the realm of intelligence education. Among these methods, knowledge tracing (KT)~\cite{kt} has emerged as a crucial task, aiming to predict students' future performance and assess their knowledge states based on their historical question responses. KT can not only improve teaching efficiency but also help students identify and address learning gaps.
To achieve effective knowledge tracing, educators must discern students' knowledge proficiency, which is not explicitly observable from their learning behaviors. The sole indicators available are students' question response records, where each response reflects their mastery level of the relevant knowledge concept. A correct response indicates higher proficiency, while an incorrect one suggests the need for improvement. Consequently, these responses serve as the foundation for KT, with prediction performance on responses serving as the gold standard.

\begin{figure}[!t]
  \centering
  \includegraphics[width=0.8\linewidth]{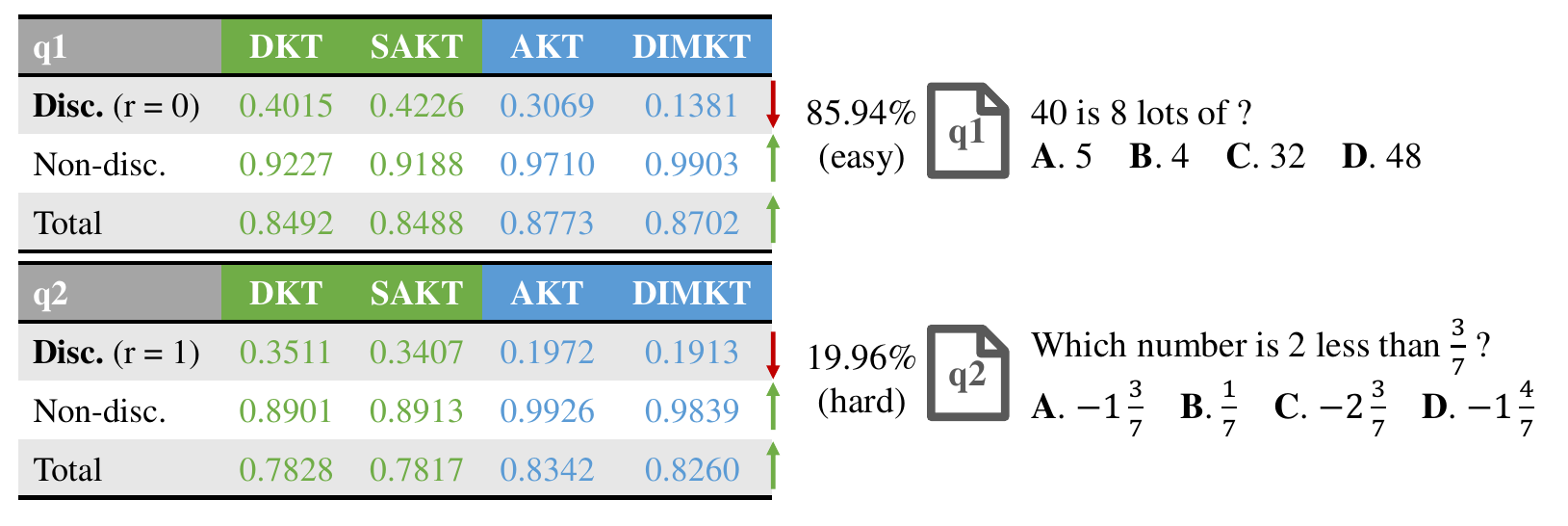}
  \caption{Two examples of questions answered by students in the Eedi dataset. The accuracy of four KT methods in predicting the responses to these questions is shown on the left.  ``Disc.'' means discrimination. $r=1$ indicates a correct response.}
 \label{fig:intro}
\end{figure}

Numerous efforts have been dedicated to the advancement of KT techniques. Bayesian knowledge tracing (BKT) ~\cite{kt} stands as an early and well-established probabilistic approach.
It serves as the foundation for various branches of KT methods~\cite{kt-idem, ibkt}.
Subsequently, the introduction of deep learning knowledge tracing approaches~\cite{dkt, sakt, akt, lpkt, mrtkt,cui2024} empowered by the capabilities of neural networks~\cite{deep-learning}, has resulted in remarkable performance improvements.
Despite these achievements, existing methods have primarily focused on enhancing prediction performance. They often disregard the discrimination imbalance present in responses.
Response discrimination gauges how effectively a question response can differentiate students with varying levels of mastery.
This discrimination depends on the question's difficulty and the binary correctness of students' responses.
To illustrate this point, Figure~\ref{fig:intro} demonstrates two examples of student answers from the Eedi dataset.
Consider $q_1$, a question with 85.94\% of students answering correctly.
Since it is easy to solve, a correct response to $q_1$ does not necessarily indicate a significant difference in the mastery of the corresponding knowledge concept, ``Mental Multiplication and Division'', compared to other students. However, an incorrect response to $q_1$ could offer better discrimination in knowledge mastery, suggesting a proficiency level lower than that of 85.94\% of the students.
Similarly, answering $q_2$ correctly can imply a high mastery level of the concept ``Adding and Subtracting Fractions''. Consequently, responses to different questions possess varying degrees of discrimination abilities. Those with high discriminative values can provide valuable insights for educators to differentiate among students and deserve more attention in KT research.

However, upon conducting real data analysis (refer to Section~\ref{sec:data-ana} for detailed findings), we have uncovered a substantial portion of weakly discriminative responses within the student response data.
This leads to an imbalanced distribution of response discrimination.
As a consequence, the current KT methods tend to disproportionately improve prediction performance on these less discriminative responses, while neglecting the predictive ability of the high discriminative ones.
The result of this trend is a deceptive surge in prediction accuracy, creating an unbalanced knowledge tracing ability.
For instance, when faced with an exceedingly challenging question that students struggle to answer correctly, all responses to this question tend to be incorrect.
Consequently, a direct prediction of these responses as incorrect would yield a high accuracy but offer minimal assistance to educators in tracing students' knowledge states and differentiating their mastery levels, which is the primary objective of KT.
As exemplified in Figure~\ref{fig:intro}, two state-of-the-art methods, AKT~\cite{akt} and DIMKT~\cite{dimkt}, exhibit enhanced accuracy in predicting responses to $q_1$ compared to their predecessors, DKT~\cite{dkt} and SAKT~\cite{sakt}.
However, this improvement is mainly attributed to their enhanced capability in predicting non-discriminative correct responses.
In other words, these advancements come at the cost of diminished performance in predicting discriminative responses to $q_1$. Similar patterns are observed in the case of $q_2$.

In the realm of KT, few works have mentioned the concept of response discrimination, let alone the imbalance issue in this context. Some works consider the question discrimination. The cognitive diagnosis approaches~\cite{irt,Liu2023} applied to KT considers question discrimination as an inherent property of questions, which is used for parameterization. However, these approaches overlooks response discrimination, neglecting the interaction with students.
On the other hand, some methods~\cite{kt-idem, dimkt} concentrate on question difficulty, which is only a question property related to response discrimination.
Previous works~\cite{dimkt, iekt} have explored the idea that answering questions with different difficulty gains different knowledge, thereby enhancing performance. While these methods consider such an idea that is similar to response discrimination, they do not explicitly address the imbalance problem that arises when handling discriminative responses in commonly-used KT methods.
This paper aims to address the imbalance issue and poses a key research question: how can we effectively tackle the discrimination imbalance issue prevalent in commonly-used KT methods? In response to this question, we propose a novel \textbf{\underline{D}iscrimination \underline{R}ebalancing framework \underline{for} \underline{K}nowledge \underline{T}racing (DR4KT)}. Our approach utilizes loss reweighting, a method that assigns different weights to guide models to focus on more important samples during loss function optimization.
Loss reweighting methods have found extensive use in various domains, such as computer vision and natural language processing~\cite{focal, dice}, due to their simplicity and versatility to address data imbalance issue~\cite{imbalance_review}.
However, applying loss reweighting to rebalance discriminative responses faces two primary challenges:
(i) The collected data lack discrimination annotations, making it difficult to effectively rebalance the data. As a result, an appropriate approach to estimate the discrimination of each response is necessary.
(ii) Rebalancing the training objective to improve performance on high discriminative responses may inadvertently reduce the original performance on low discriminative responses.
This is undesirable and the performance should be preserved in an effective solution.

To address the first challenge, we introduce a well-defined numerical measure of response discrimination that quantifies a question response's ability to differentiate students effectively.
However, directly computing discrimination scores using statistics from the collected data encounters a sparsity issue.
This is particularly apparent when some questions are answered by only a few students, leading to noisy and inaccurate discrimination scores.
To mitigate this problem, we propose a frequency-aware question correctness tendency estimator. This estimator takes into account the information about a given question and its occurrence frequency.
The output of this estimator is a question correctness tendency score, which represents the probability that a random student would answer the question correctly. This score captures the overall tendency of the entire student group towards answering the question, thus reflecting the question's difficulty. A higher difficulty indicates a lower tendency of students to answer the question correctly.
Subsequently, we combine the question correctness tendency score with the ground-truth correctness label of a specific response (i.e., 1 for correct answers and 0 for incorrect answers) to generate a response weight. This response weight is used to reweight the loss during model training. By using this approach, we can effectively assign different weights to responses based on their discrimination levels.
It mitigates the impact of sparsity in the data and enables more accurate loss reweighting during training.


To address the second challenge of maintaining high prediction accuracy on less discriminative responses, we introduce an adaptive predictive score fuser.
This fuser finalizes predictions according to a discrimination-aware rule, assigning primary responsibility to the reweighted KT model for predicting more discriminative responses. Simultaneously, the prediction for less discriminative responses is complemented by the question correctness tendency estimator.
This approach is motivated by the understanding that less discriminative responses offer limited information for distinguishing among students.
Consequently, their response correctness tends to align with the overall student group, which could be approximated by the question correctness tendency. 
However, during inference, the response discrimination score is unknown as it stems from the response correctness.
To address this, we predict the response discrimination score using a Multilayer Perceptron (MLP).
The MLP is fed with knowledge state vectors extracted from the original KT model and informative question representations learned by the question correctness tendency estimator.
Subsequently, we align the output with the ground-truth response discrimination score in a joint task learned alongside the overall optimization.

Through a suitable combination of the reweighted KT model and the question correctness tendency estimator, we achieve significant performance improvements on mainstream deep learning-based KT methods. This demonstrates the effectiveness of our DR4KT framework in tackling the imbalance issue and enhancing the overall KT performance.
To the best of our knowledge, DR4KT is the first method that addresses the discrimination imbalance of student responses in the context of KT. The key contributions of our work can be summarized as follows:
\begin{itemize}[leftmargin=*]
\item \textbf{Discovery.} Through thorough data analysis on real datasets, we have identified and brought attention to the presence of imbalanced discrimination among student responses in KT scenarios. Moreover, we have highlighted the significant impact of this issue, as it leads to meaningless inflation of prediction accuracy and hampers the effectiveness of knowledge tracing.

\item \textbf{Method.} Our proposed DR4KT framework presents a model-agnostic approach to rebalance discriminative responses, thereby improving KT models' overall knowledge tracing ability. Importantly, our method ensures that the contribution from less discriminative responses to prediction accuracy is preserved.
This mitigates the potential decline in accuracy of such responses.

\item \textbf{Experiments.} We conducted extensive experiments on three widely-used datasets to evaluate the effectiveness of DR4KT when applied to several typical KT methods. The experimental results demonstrate significant improvements in knowledge tracing performance, thus validating the efficacy of our proposed approach.
\end{itemize}

\section{background}

\subsection{Learning from Imbalanced Data}

Imbalanced data distributions are prevalent in specific domains, such as fraud detection and cancer diagnosis. 
In these areas, the abundance of majority samples tends to overshadow the training process, leading to suboptimal performance in predicting crucial yet rare minority samples.
To address this challenge and achieve favorable outcomes, researchers recommend employing sampling-based and cost-sensitive methods~\cite{imbalance_review}.
Specifically, the first category aims for balance through oversampling minority samples or undersampling majority samples.
A noteworthy oversampling method is SMOTE~\cite{smote}, which linearly combines minority samples with their neighboring counterparts to generate new minority samples.
Building upon this method, subsequent approaches have emerged, demonstrating commendable performance~\cite{han_smote,douzas_smote}.
In addition, certain undersampling methods randomly select an equal number of informative majority samples for minority samples multiple times to train multiple balanced models.
Subsequently, bagging or boosting techniques are applied to ensemble these models~\cite{undersampling,sun_undersampling,zhang_undersampling}.

Another category, cost-sensitive methods, directs attention to assigning varying costs to models when they misclassify different samples during the learning stage.
This learning-oriented approach enhances efficiency in the face of the current data explosion, making it a widely adopted strategy for addressing imbalanced data.
Some rule-based machine learning methods incorporate cost matrices to prioritize the decision rules that are more relevant for the minority class~\cite{lopez2015cost,cost_tree,gupta2022cse}.
For instance, Krawczyk \textit{et al.}~\cite{cost_tree} proposed an evolutionary algorithm to select the decision tree ensembles that have the lowest misclassification cost.
Besides, some cost-sensitive methods reweight objective functions to guide models to focus on minority samples.
The work~\cite{maldonado2014imbalanced} directly uses margin variables associated with each class while optimizing SVMs.
Lin \textit{et al.}~\cite{focal} introduced focal loss, for dense object detection in computer vision. This method adapts the loss value according to the difficulty level of each sample’s classification, and has been proven to be effective in addressing the long-tailed imbalance problem, which arises when some classes are much less frequent than others.
We note that our DR4KT adopts the loss reweighting scheme, which belongs to this category, and focuses on the imbalance of response discrimination in KT.

\subsection{Loss Reweighting}

loss reweighting has proven to be a versatile and effective technique used in various fields to address a range of issues, particularly in handling class imbalance. Its capability to assign different weights to individual samples makes it a powerful tool for resolving such challenges.
In the domain of computer vision, Lin \textit{et al.}~\cite{focal} introduced focal loss for dense object detection.
It focuses on the classification difficulty of each sample. This approach has significantly contributed to tackling the long-tailed problem in computer vision tasks~\cite{class-balance, ghm}, where some classes are heavily underrepresented.
Additionally, in the context of classification tasks, the work~\cite{vnet} proposed DICE loss, which directly optimizes the F1-score for classification, thereby providing a loss reweighting scheme. Inspired by this, DSC loss~\cite{dice} adapted the concept of loss reweighting to natural language processing (NLP) tasks, achieving notable performance improvements in various NLP applications.

In the realm of recommender systems, loss reweighting is a popular technique due to its efficiency and ability to generalize across various scenario requirements and large datasets. Recommender systems often need to handle vast amounts of data and adapt to diverse user preferences and interactions.
For instance, YouTube~\cite{youtube} utilizes loss reweighting to balance the objective function based on the video-watching time of their users. This approach allows the recommender system to take into account the varying levels of engagement with different videos.
Another significant application of loss reweighting in this field is to address multiple system biases using Inverse Propensity Weighting (IPW) methods~\cite{ipw}. IPW-based methods involve multiplying an inverse propensity score by each sample's loss term.
They effectively reweight the observed data to approximate the underlying real unbiased data distribution~\cite{rec-debias, position}. By doing so, these methods aim to mitigate biases present in the recommender system and provide fair and accurate recommendations to users.

In KT, some existing studies have focused on improving knowledge tracing by reconstructing loss functions. For instance, Chen~\textit{et al.}~\cite{pdkt} proposed the partial-order loss, which incorporates prerequisite knowledge concept relations to enhance the knowledge tracing process. Additionally, Lee~\textit{et al.}~\cite{clkt} effectively employed contrastive learning in KT, achieving strong performance in knowledge tracing tasks.
However, these approaches typically treat each response equally during optimization, without considering the importance of individual responses to the knowledge tracing process. As a result, there is limited research on effectively measuring the significance of each response in the KT framework.
Although directly applying loss reweighting to KT seems like a straightforward solution, it faces challenges, such as the question sparsity problem, and may lead to performance declines in predicting responses with small weights. In this paper, we handle this issue in DR4KT.

\subsection{Knowledge Tracing}

Knowledge tracing~\cite{kt} has been a long-standing research topic and has seen the development of various types of methods. Among them, probabilistic models based on BKT~\cite{kt} are prominent examples. These models adopt a Hidden Markov Model (HMM) framework to sequentially predict students' future responses by leveraging transition and emission probabilities.

The recent advancements in deep learning have indeed proven to be beneficial for KT
They offer a wealth of opportunities to improve the KT performance from various perspectives. 
One of the representative methods that emerged in the context of KT is Deep Knowledge Tracing (DKT)~\cite{dkt}. DKT is designed to sequentially model students' historical responses, allowing for a more comprehensive understanding of their knowledge states over time. Due to its effectiveness, DKT has served as a foundation for subsequent approaches in KT, leading to the development of methods~\cite{dkt-forget,dbn,lpkt,dimkt,ekt}.
Furthermore, the integration of attention mechanisms~\cite{attention} into KT methods has proved to be valuable. Attention mechanisms enable the capture of each historical response's contribution to the correct answering of a new question. This allows the model to focus on relevant historical responses and effectively adapt to each student's learning trajectory~\cite{sakt,akt,rkt,ekt}.

Moreover, side information has been leveraged in several methods to KT. 
HawkesKT~\cite{hawkesKT} is a notable example of a method that utilizes temporal information. It employs a Hawkes process-based approach to capture temporal cross effects between historical questions and a target question.
Such temporal information is also utilized by LPKT~\cite{lpkt}.
It considers the consistency of the learning and forgetting process over time and takes into account the temporal patterns of students' responses to different questions.
The study~\cite{huang2020learning} effectively leverages the learning and forgetting curves to model student learning behaviors.
Furthermore, some studies~\cite{ekt, rkt} have integrated textual information from questions to enhance their embeddings.
By incorporating text-based features, these methods can better represent the content and context of questions, leading to more informative and effective knowledge tracing models.
Recently, some studies~\cite{counterfactual-kt,shen2024} pursue interpretable knowledge tracing that explains the decision process of predicting student performance on target exercises.

This paper focuses on resolving response discrimination imbalance in KT, which is orthogonal to existing KT methods and could incorporate them into the proposed DR4KT framework.


\subsection{Response Discrimination}
Response discrimination in KT measures how effectively a student's response to a question can distinguish their knowledge mastery level.
Different from question discrimination, it considers both question difficulty and response correctness.
In KT, these crucial factors---question difficulty, question discrimination, and response discrimination---are essential for accurately assessing knowledge proficiency from a psychometric perspective~\cite{dimkt, kt-idem, iekt}.
We use 2PL-IRT \cite{irt} to depict their distinctions and connections.
The probability of a student answering a question $i$ correctly is derived from a logistic function
\begin{equation}\label{eq:irt}
p_i(\theta)=\frac{1}{1+e^{-a_i(\theta-b_i)}},
\end{equation}
where $\theta$ is the measured proficiency, and $a_i$ and $b_i$ respectively denotes the question discrimination and difficulty parameters.
Observations from previous works~\cite{item-diff-disc, norming} suggest that questions with a moderate difficulty level are associated with high discrimination; that is, questions that are too easy or too hard are challenging to distinguish students' knowledge states. To provide a brief proof, consider two students $j$ and $k$ with proficiency levels $\theta_j$ and $\theta_k$. The discrimination level for question $i$ can be reflected by the joint probability of the more proficient student of them answering $i$ correctly and the less proficient student answering $i$ incorrectly. This probability reaches its maximum when the difficulty parameter $b_i$ equals to $(\theta_j+\theta_k)/2$ and decreases when $b_i$ deviates from this value. The parameter $a_i$ controls the rate of decrease.

However, when combined with student responses, easier and harder questions might exhibit more discrimination than moderate ones.
Also giving $\theta_j,\theta_k$, and an extra condition that a student (e.g., $j$) responds correctly, the probability to discriminate two of them, i.e., the more proficient student answering $i$ correctly and the less proficient student answering $i$ incorrectly, instead becomes $(1-p_i(\theta_k))/2$, which increases monotonically with $b_i$.
If the condition is changed to an incorrect response, the probability is then $p_i(\theta_k)/2$ and decreases monotonically vice versa.
This observation aligns with common sense, as a correct answer to a hard question indicates high knowledge mastery, and an incorrect answer to an easy question indicates low knowledge proficiency. Some works have leveraged this phenomenon to improve KT performance~\cite{dimkt, iekt}.
We, however, regard this as response discrimination and address its imbalance issue in KT.
To our best knowledge, this is the first study to explore and tackle the issue of response discrimination imbalance in the KT field.

\section{Preliminary}
In this section, we will start by providing a clear definition of the KT tasks. Subsequently, we will present a formal description of response discrimination, and highlight its significance in the context of KT. Finally, we will conduct a comprehensive analysis using real data and various KT methods to demonstrate the detrimental consequences of disregarding discriminative responses.

\subsection{Knowledge Tracing}

Knowledge tracing, which traces students' knowledge mastery by predicting their performance on given questions, follows the setup below.
Suppose we have a student set $\Uset$, a question set $\Qset$, and a knowledge concept set $
\mathcal{C}$.
Denote $r^u_i=(q^u_i, a^u_i, \Kset_i^u)$ as the $i^{\text{th}}$ response of student $u\in\Uset$. 
It consists of the responded question $q_i^u$, its related knowledge concepts $\Kset_i^u\subset\mathcal{C}$ and the binary response correctness $a^u_i\in\{0, 1\}$, with $a^u_{i}$=1 when the response is correct.
Among them, $\Kset_i^u=\{k^u_{i,1},k^u_{i,2},\cdots,k^u_{i,|\Kset_i^u|}\}$ represents the concept set related to the question $q_i^u$, where $k^u_{i,j}$ is its $j^{\text{th}}$ concept.
Given the $t$ historical responses of the student as a sequence $\Hset^u_t=\{r^u_{1},r^u_{2},\cdots,$ $r^u_{t}\}$, knowledge tracing aims to know whether $u$ could answer a new assigned question $q^u_{t+1}$ correctly, which is denoted as $a^u_{t+1}$.
As such, any KT model could be represented as a predictive function to derive the probability score to correctly answer the target question:
\begin{equation}
    \label{eq:kt-pre}
    \hat{a}^u_{t+1}=\psi(q_{t+1}^u,\Kset_{t+1}^u, \Hset^u_t|\boldsymbol{\Theta}_\psi)
\end{equation}
where $\boldsymbol{\Theta}_\psi$ is the model parameters. 
It is worth noting that most mainstream KT approaches~\cite{akt, lpkt, dimkt, iekt} adopt the KT definition that involves both question and concept information, which we also do in our setting.
Moreover, some methods~\cite{hawkesKT,lpkt,ekt} leverage side information. However, in this study, we focus on a general framework to improve KT performance and omit the details of these side information.
Besides, we summarize the key mathematics notations of DR4KT in Table~\ref{tab:notations} to illustrate the model structure and inference procedure clearly.
The bold upper case letters denote matrices and bold lower case letters denote vectors.

\begin{table}[!t]
\centering
\caption{Key mathematical notations.}
\label{tab:notations}
\begin{tabular}{l|l}
    \hline
    Notations & \multicolumn{1}{c}{Description}\\
    \hline
    $\Uset,\Qset,\Cset$ & student, question and concept sets \\
    $u, \Hset_t^{u}$ & student, and its $t$-length historical response sequence \\
    $t,T_u$ & historical response length, and historical response length of $u$\\
    $r,r_{i}^{u}$ & response, and $i^{\text{th}}$ response of student $u$ \\
    $q,q_{i}^{u}$ & question, and question of response $r_{i}^{u}$ \\
    $\Kset,\Kset_{i}^{u}$ & knowledge concept set, and knowledge concept set of response $r_{i}^{u}$\\
    $a,a_{i}^{u}$ & response correctness, and response correctness in response $r_{i}^{u}$ \\
    $\psi(\cdot),\boldsymbol{\Theta}_{\psi}$ & KT model and its network parameters\\
    $\hat{a}^{u}_{t+1}$ & predicted probability of student $u$ correctly answering $q^u_{t+1}$ by KT models\\
    $b,b_i^u$ & correctness tendency, and correctness tendency of $q_i^u$\\
    $\tilde{b}$ & estimated correctness tendency in data analysis\\
    $\hat{b}$ & estimated correctness tendency in DR4KT\\
    $\delta_i^u$ & response discrimination of $r_i^u$\\
    \hline
    $d$ & number of hidden dimensions\\
    $\textbf{q},\textbf{k},\textbf{f}$ & question, concept and frequency embeddings\\
    $\textbf{e},\textbf{e}_i^u$ & question representation, and question representation of $q_i^u$\\
    $\textbf{m}_i^u$ & extracted knowledge state vector from KT backbone when answering $r_i^u$\\
    $\phi(\cdot),\boldsymbol{\Theta}_{\phi}$ & frequency-aware question correctness tendency estimator and its parameters\\
    $\textbf{w}^{\textbf{T}}_{\phi},\beta$ & network parameters in correctness tendency estimator\\
    $\hat{b}_i^u$ & estimated correctness tendency of $q_i^u$ in DR4KT\\
    $\hat{\delta}_i^u$ & estimated response discrimination of $r_i^u$ in DR4KT in training\\
    $w_i^u$ & discrimination-aware loss weight of response $r_i^u$\\
    $\kappa(\cdot),\boldsymbol{\Theta}_{\kappa}$ & MLP in adaptive predictive score fusion and its parameters\\
    $\textbf{W}_{\kappa}^1, \textbf{W}_{\kappa}^2, \textbf{b}_{\kappa}^1, \textbf{b}_{\kappa}^2$ & network parameters in MLP in adaptive predictive score fusion\\
    $\tilde{\delta}_i^u$ & predicted response discrimination of $r_i^u$ in DR4KT in inference\\
    $\xi_i^u$ & discrimination-aware score fuser of response $r_i^u$\\
    $\hat{y}^{u}_{i}$ & fused probability score of correctly answering $q^u_{i}$ by DR4KT\\
    \hline
    $\tau_1,\tau_2$ & hyper-parameters to control the weight and fuser transformation \\
    $\lambda_1,\lambda_2$ & hyper-parameters to balance loss re-weighting and discrimination alignment\\
    \hline
    
\end{tabular}
\end{table}

\subsection{Response Discrimination}

Response discrimination is a crucial aspect in KT as it measures a response's ability to distinguish a student's knowledge mastery level from other students. It takes into account both the difficulty of the question and the student's response to quantify this ability accurately.
To quantify this ability, we first introduce question difficulty. Question difficulty indicates the effort or skill required to solve a question. We follow a common setting that uses the probability of students answering a question correctly to reflect its difficulty~\cite{diff-rate,dimkt}, and we define this probability as the question's \textbf{correctness tendency}.
The higher the difficulty, the lower the correctness tendency.
Given a question $q$ and its knowledge concepts $\Kset$, its correctness tendency is the probability of its response correctness $a = 1$, which is
\begin{equation}\label{eq:difficulty}
    b = p(a=1|q,\Kset).
\end{equation}
Now, we can quantify the response discrimination, which reflects how well a response can differentiate a student from others. To achieve this, we define the discrimination of response $r_i^u$ as the probability of other students not getting the same response correctness as $r_i^u$. We express this probability using the correctness tendency as
\begin{equation}\label{eq:disc_prob}
\delta^u_i = \left\{\begin{aligned}
&b^u_i,\quad&a^u_i = 0\\
&1-b^u_i,\quad&a^u_i = 1
\end{aligned}
\right.,
\end{equation}
where $b_i^u$ is the correctness tendency of $q_i^u$.
This definition aligns with our intuition that correct responses to difficult questions (low correctness tendency) are more discriminative, as well as incorrect responses to easy questions.

\begin{figure*}[t]
  \centering
  \includegraphics[width=\linewidth]{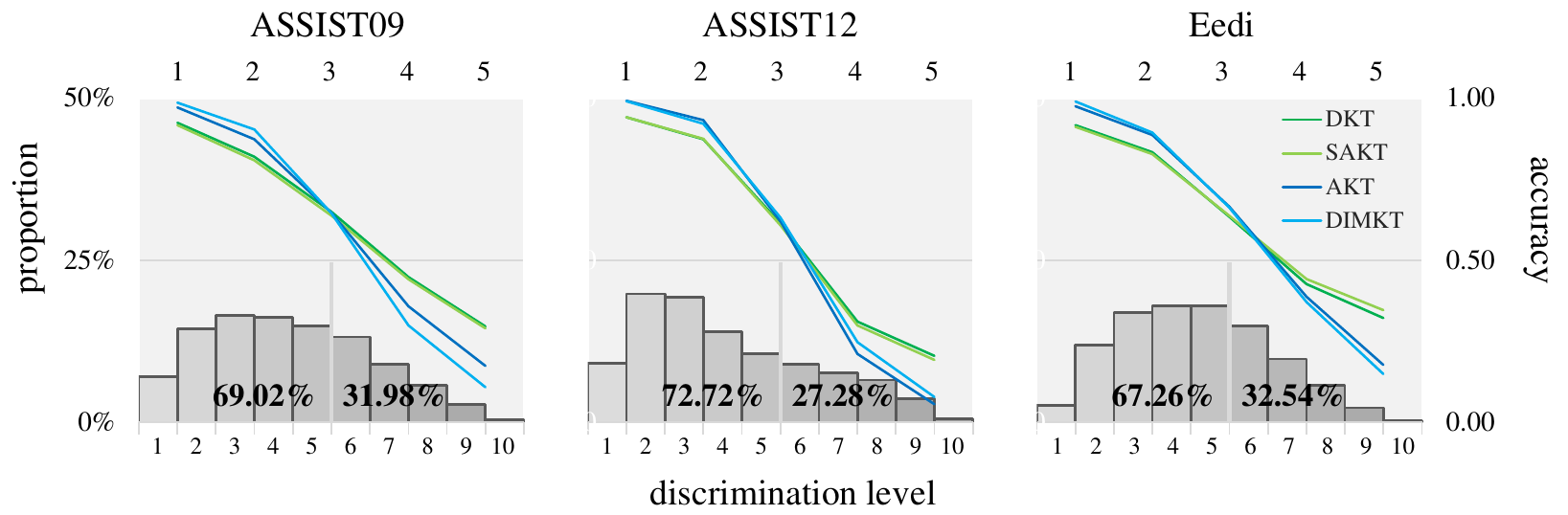}
  \caption{Response proportions and prediction accuracy of four typical KT methods at different discrimination levels of three datasets. The percentage pairs indicate the proportions of low discriminative responses (from 0 to 0.5) and high ones (from 0.5 to 1).}
 \label{fig:disc}
\end{figure*}

\begin{table}[t]
\begin{center}
\caption{Prediction accuracy of four typical KT methods on three datasets.}
\begin{tabular}{l|ccc|l|ccc}
\hline\hline
Model & ASSIST09 & ASSIST12 & Eedi   & Model & ASSIST09 & ASSIST12 & Eedi   \\ \hline
DKT   & 0.7248   & 0.7345   & 0.7049 & AKT   & 0.7344   & 0.7490   & 0.7325 \\
SAKT  & 0.7156   & 0.7314   & 0.7061 & DIMKT & 0.7351   & 0.7531   & 0.7330 \\ \hline\hline
\end{tabular}
\label{tab:pre_acc}
\end{center}
\end{table}

\begin{table}[t]
\setlength{\tabcolsep}{3.2pt}

\begin{center}
\caption{Prediction accuracy of DKT after dropping varied proportions of highest and lowest discriminative responses during inference. The cross term of ``-Highest''/``-Lowest'' and x\% indicates the performance after we remove the highest/lowest x\% of discriminative responses in each dataset.}
\begin{tabular}{l|cccc|cccc|cccc}
\hline\hline
Dataset   & \multicolumn{4}{c|}{ASSIST09}                   & \multicolumn{4}{c|}{ASSIST12}                   & \multicolumn{4}{c}{Eedi}                        \\ \hline
Proportion  & 5\% & 10\% & 15\% & 20\%  & 5\% & 10\% & 15\% & 20\% & 5\% & 10\% & 15\% & 20\%\\\hline
-Highest & 0.7177 & 0.7094 & 0.7015 & 0.6933 & 0.7322 & 0.7298 & 0.7253 & 0.7218 & 0.6898 & 0.6818 & 0.6742 & 0.6623 \\
-Lowest & 0.7225 & 0.7204 & 0.7169 & 0.7126 & 0.7332 & 0.7316 & 0.7290 & 0.7250 & 0.6995 & 0.6907 & 0.6797 & 0.6675 \\ 
\hline\hline
\end{tabular}
\label{tab:drop_acc}
\end{center}
\end{table}

\subsection{Data Analysis}\label{sec:data-ana}
In this part, we present the results of our real data analysis conducted on three widely-used datasets, as described in Section~\ref{subsec:data}. The analysis has led to two main discoveries:
\begin{itemize}[leftmargin=*]
\item KT scenarios suffer from an imbalance of response discrimination. This means that there is a prevalence of low discriminative responses in the datasets.
\item Improved performance of state-of-the-art methods is primarily achieved by predicting low discriminative responses more accurately. In contrast, the performance of high discriminative responses tends to deteriorate.
\end{itemize}

In order to estimate response discrimination, we approximate the correctness tendency of a question $q$ in Equation~\ref{eq:difficulty} using the question passing rate, denoted as $\tilde{b}$, which is
\begin{equation}
\tilde{b}=\frac{\sum_{u\in\Uset}\sum_i^{T_u}I\left(a_i^u=1\right)I\left(q_i^u=q\right)}{\sum_{u\in\Uset}\sum_i^{T_u}I\left(q_i^u=q\right)}
\end{equation}
where $I(\cdot)$ is the indicator function that takes the value 1 if the input Boolean expression is true.
$T_u$ denotes the length of response sequence of student $u$.
We then calculate each response's discrimination score denoted as $\tilde{\delta}_i^u$ by using the same conversion as in Equation~\ref{eq:disc_prob}.
After that, we divide all responses according to ten uniform discrimination levels (i.e., 0.1 per interval) and display their proportions in Figure~\ref{fig:disc}.
In the analysis, responses related to questions answered less than ten times are omitted to exclude the problem of question sparsity and resultant noisy estimation in this approximation.
As shown, low discriminative responses (0.0-0.5) account for a large portion of nearly 70\% in all three datasets, while high discriminative ones (0.5-1.0) are scarse.
In fact, high discriminative responses have a strong ability to distinguish students' knowledge mastery, and thus merit more attention.
Current methods only enhance the overall prediction accuracy without considering the imbalanced distribution of discrimination, thus impairing the knowledge tracing ability.
To illustrate this phenomenon, we train four commonly-used KT methods: DKT, SAKT, AKT, and DIMKT~\cite{dkt,sakt,akt,dimkt}.
DKT and SAKT represent earlier methods, while AKT and DIMKT represent the state-of-the-art methods.
The performance of the two newer methods is shown in Table~\ref{tab:pre_acc}, and they demonstrate an improvement compared to the older ones.
We also evaluate their performance on responses with five discrimination levels (0.2 per interval) in Figure~\ref{fig:disc}.
The results reveal that while AKT and DIMKT achieve higher overall prediction accuracy compared to DKT and SAKT.
They are less effective in predicting high discriminative responses.
This suggests that the accuracy improvement is primarily driven by lower discriminative responses, while high discriminative responses are not given enough attention, leading to less meaningful improvement in knowledge tracing.

Furthermore, we conduct another experiment to validate the importance of high discriminative responses.
We gradually drop a certain proportion of highest and lowest discriminative responses during knowledge tracing inference, and then compare their performance shift before and after the dropping.
We choose the typical method DKT as the experimental method.
The results are shown in Table~\ref{tab:drop_acc}.
As can be seen, after dropping the same proportion of responses, the performance shift of the high discriminative response dropping is larger than the shift of dropping low discriminative responses.
Such shift gap exactly indicates the key role of high discriminative responses contributing to trace student knowledge states.

Motivated by these observations, we propose a general loss reweighting framework to encourage KT models to prioritize more discriminative responses.
It promotes knowledge tracing performance and addressing the imbalance issue of response discrimination.

\section{methodology}

\begin{figure*}[!t]
  \centering
  \includegraphics[width=0.9\linewidth]{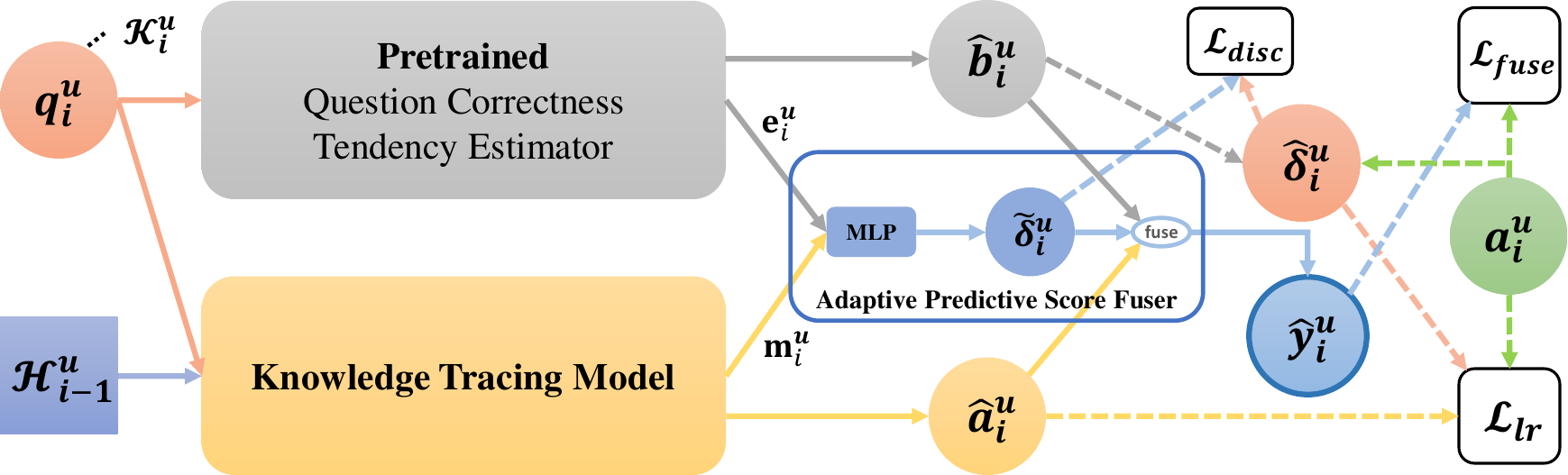}
  \caption{The entire framework of DR4KT. The question correctness tendency estimator is pretrained in advance. The dashed arrows indicate the inference only in training.}
 \label{fig:dr4kt}
\end{figure*}

This section elaborates on our DR4KT framework. 
We first introduce a pretrained frequency-aware question correctness tendency estimator, which provides question correctness tendency scores as the measurement of question difficulty.
Then, we estimate the discrimination score of each response using the obtained correctness tendency score while training the KT model.
Based on this, we apply loss reweighting and an adaptive predictive score fuser to prepare for the final student performance prediction.
The whole framework is presented in Figure~\ref{fig:dr4kt}.

\subsection{Frequency-aware Question Correctness Tendency Estimator}

As mentioned before, we use question correctness tendency to indicate question difficulty. A lower difficulty suggests a higher correctness tendency.
To estimate this question correctness tendency as in Equation~\ref{eq:difficulty}, we propose a frequency-aware question correctness tendency estimator.

A direct method to estimate the correctness tendency is using the passing rate of questions in the training data, but it can encounter data sparsity issues, as many questions are answered only a few times.
We resort to using knowledge concepts which are non-sparse and also provide difficulty information from the concept aspect.
To obtain a generalized and noiseless correctness tendency score $\hat{b}$ of a question $q$, we employ a frequency-aware embedding fuser and a fully-connected network that deeply fetch questions' inherent representations.
It is denoted as a network $\hat{b}=\phi(q|\boldsymbol{\Theta}_\phi)$, where $\boldsymbol{\Theta}_\phi$ is the learnable parameters.
For cold questions with low frequencies, we expect that more difficulty information comes from the knowledge concept associated with the question.
To address this, we count and sort the frequencies of questions and split them into $N$ groups, ensuring that each group has the same sum of question frequencies. We then assign a frequency embedding to each question based on the group it belongs to.
Thus, we derive the representation of the question $q$ by a frequency-aware fusion gate:
\begin{equation}
    \label{eq:question}
\textbf{e}=\left(\frac{1}{|\Kset|}\sum_{k\in\Kset}\textbf{k}\right)\circ\sigma\left(\textbf{f}\right) + \textbf{q}\circ(1-\sigma(\textbf{f})),
\end{equation}
where $\textbf{q}\in\mathbb{R}^{d}$ and $\textbf{k}\in\mathbb{R}^{d}$ are the ID embeddings of the question $q$ and its one related concept $k$. $\textbf{f}\in\mathbb{R}^{d}$ is its frequency embedding and $\circ$ is the Hardamard product.
$d$ is the embedding dimension number and $\sigma(\cdot)$ is the sigmoid function.
Then, we attain the final correctness tendency score by a fully-connected network
\begin{equation}
    \label{eq:diff_score}
\hat{b} = \sigma\left(\textbf{w}^{\textbf{T}}_\phi\textbf{e}+\beta\right),
\end{equation}
where $\textbf{w}_\phi\in\mathbb{R}^d$ and $\beta\in\mathbb{R}$ are the network parameters.

Henceforward, the correctness tendency score of each question is obtained and then leveraged by the subsequent modules of DR4KT.
Besides, this correctness tendency estimator is pretrained, and its learning process will be further explained later.
It is worth noting that, there are two cases not suitable for this frequency-aware fusion.
One case is for the completely new questions with no historical response records.
As an alternative, we could assume their frequencies as 0 and directly use the embeddings of their related concepts.
Another case is that few scenarios do not provide both problem and concept annotation of each response.
For this, we could directly use the problem or concept embeddings instead.
Both these cases are rare in the KT scenarios and thus do not affect the generality of DR4KT.
The proposed alternatives are also compatible with the subsequent process of DR4KT.

\subsection{Rebalancing Discriminative Responses}
In KT tasks, the usual training objective is to optimize the binary cross-entropy loss between the predicted response correctness $\hat{a}^u_{i}$ and the ground-truth $a^u_{i}$:
\begin{equation}
    \label{eq:bce}
    \mathcal{L}_{kt}=-\sum_{u}\sum_{i}^{T_u}\left(a^u_i\text{log}\hat{a}^u_i+(1-a^u_i)\text{log}(1-\hat{a}^u_i)\right).
\end{equation}
For simplicity, we omit the average operation notation, and we do the same hereafter.

However, as we have analyzed before, KT models should focus more on discriminative responses to improve knowledge tracing quality. To achieve this, we employ a loss reweighting technique that assigns each response's loss term a new weight $w_i^u$, which is formulated as
\begin{equation}
    \label{eq:lr-bce}
    \mathcal{L}_{lr}=-\sum_{u}\sum_{i}^{T_u}w^u_i\left(a^u_i\text{log}\hat{a}^u_i+(1-a^u_i)\text{log}(1-\hat{a}^u_i)\right).
\end{equation}
This weight is discrimination-aware and gives high weights to discriminative responses.
Therefore, we first extract the correctness tendency score $\hat{b}_i^u$ of $q_i^u$ from the correctness tendency estimator and then obtain the discrimination score of $r_i^u$ by reformulating Equation~\ref{eq:disc_prob} as
\begin{equation}
    \label{eq:rewrite}
\hat{\delta}_i^u = a_i^u(1-\hat{b}_i^u)+\left(1 - a_i^u\right)\hat{b}_i^u.
\end{equation}
Afterwards, we calculate the assigned weight by:
\begin{equation}
    \label{eq:d2w}
    w_i^u = e^{\text{log}(\hat{\delta}^u_i)/\tau_1},
\end{equation}
where $\tau_1$ is a hyper-parameter to control the intensity of reweighting that a higher value means less rebalancing.
By using this design, the discrimination scores are appropriately transformed to weights for each response without changing their value intervals (from 0 to 1). This approach is beneficial for the quantization and tuning of response contribution during the training process, which ensures that higher discriminative responses are given more attention in the knowledge tracing model.

\subsection{Adaptive Predictive Score Fusion}
To address the degradation in prediction performance on low discriminative responses caused by loss reweighting, we propose an adaptive trade-off approach to make the final prediction.
This approach involves a fusion between the prediction score from the reweighted KT models and the score from the question correctness tendency estimator.
The idea behind this adaptive trade-off is to leverage the strengths of both these two types of scores, based on a discrimination-aware rule:
The reweighted KT models are better at handling higher discriminative responses, which are directly influenced by students' individual characteristics. On the other hand, the question correctness tendency scores reflect the overall correctness tendency of all the students answering questions and are more suitable for lower discriminative responses.
Since the discrimination score of each response relies on its correctness label, which is unknown during inference, we employ an MLP $\tilde{\delta} = \kappa(\cdot|\boldsymbol{\Theta}_\kappa)$ to predict the discrimination score by giving the student hidden knowledge state and the informative question representation.
This is formulated as
\begin{equation}
    \label{eq:disc_gen}
    \tilde{\delta}_i^u = \text{MLP}\left([\textbf{m}_i^u\oplus\textbf{e}^u_i]\right).
\end{equation}
The MLP is a two-layer feed-forward network
\begin{equation}
    \label{eq:mlp}
    \text{MLP}(\textbf{z})= \sigma\left(\textbf{W}^2_\kappa\text{ReLU}\left(\textbf{W}^1_\kappa\textbf{z}^\textbf{T}+\textbf{b}^1_\kappa\right)+\textbf{b}^2_\kappa\right),
\end{equation}
where $\textbf{W}^1_\kappa\in\mathbb{R}^{d\times 2d}, \textbf{W}^2_\kappa\in\mathbb{R}^{1 \times d}, \textbf{b}^1_\kappa\in\mathbb{R}^{d\times1}$, and $\textbf{b}^2_\kappa\in\mathbb{R}^{1\times1}$ are the model parameters to be trained.
$\textbf{e}_i^u$ is the question representation from the correctness tendency estimator (Equation~\ref{eq:question}).
$\textbf{m}_i^u\in\mathbb{R}^{d}$ is the extracted knowledge state vector from the KT model.
It contains rich personalized information about students' current knowledge states and is easy to fetch in practice.
Taking DKT as an example, its knowledge state vectors are the hidden vectors derived from RNN.
In addition, $d$ is the hidden dimension number, the same as the embedding dimension we set. $\oplus$ is the concatenation operator.
Based on this discrimination score, we perform an adaptive fusion between the KT model's score, representing the student's personalized knowledge state, and the correctness tendency score, representing the question's inherent correctness tendency:
\begin{equation}
    \label{eq:fuse}
    \hat{y}_i^u = \xi_i^u\hat{a}_i^u + (1 - \xi_i^u)\hat{b}_i^u,
\end{equation}
where $\xi_i^u$ is transformed by
\begin{equation}
    \label{eq:final-pre_3}
    \xi_i^u = e^{\log(\tilde{\delta}_i^u)/\tau_2}.
\end{equation}
The hyper-parameter $\tau_2$ controls the balance that a less value indicates a more involvement of the KT model's score.
In this way, the discrimination-aware rule is established: A higher discrimination score $\hat{\delta}_i^u$ brings a greater value of the fuser $\xi_{i}^u$, which makes the score $\hat{a}^u_i$ from the re-weighted KT model more involved into the final prediction score.
Contrarily, less $\hat{\delta}_i^u$ makes the final score closed to the overall question correctness tendency $\hat{b}_i^u$.

In this way, we could obtain the final prediction while not sacrificing the performance on less discriminative responses.
Such score fusion involves the output scores and hidden knowledge state vectors from the re-weighted KT model, constituting the core component of DR4KT extending KT models.

\subsection{Model Training}
The training process for the DR4KT framework involves pretraining the question correctness tendency estimator and then performing a joint training of the reweighted KT model and the adaptive prediction score fuser.
In the pretraining step, we train the question correctness tendency estimator on the training set using the cross-entropy loss
\begin{equation}
    \label{eq:pretrain}
    \mathcal{L}_{pre} = -\sum_{u}\sum_i^{T_u}\left(a_i^u\text{log}\hat{b}_i^u+(1-a_i^u)\text{log}(1-\hat{b}_i^u)\right).
\end{equation}
This step captures the overall response tendency of the entire student group answering each question, which provides the correctness tendency scores for them.
The frequency-aware fusion with knowledge concepts helps alleviate the question sparsity issue. 
After this, we jointly train DR4KT along with reweighting the KT model and aligning the predicted discrimination score from Equation~\ref{eq:disc_gen} with its ground-truth from Equation~\ref{eq:rewrite}.
This alignment is performed as a Mean Square Error (MSE) loss function
\begin{equation}
    \label{eq:align}
    \mathcal{L}_{disc} = \sum_{u}\sum_i^{T_u}\left(\tilde{\delta}_i^u-\hat{\delta}_i^u\right)^2.
\end{equation}
Then the joint training is conducted as
\begin{equation}
    \label{eq:joint}
    \mathcal{L} = \mathcal{L}_{fuse} + \lambda_1\mathcal{L}_{lr} + \lambda_2\mathcal{L}_{disc},
\end{equation}
where $\lambda_1$ and $\lambda_2$ are hyper-parameters to control the importance of the two auxiliary tasks.
$\mathcal{L}_{fuse}$ is the main KT loss function for the fused predicted scores $\hat{y}^u_i$ with the ground-truth response correctness $a^u_i$:
\begin{equation}
    \label{eq:main_loss}
    \mathcal{L}_{fuse}=-\sum_{u}\sum_{i}^{T_u}\left(a^u_i\text{log}\hat{y}^u_i+(1-a^u_i)\text{log}(1-\hat{y}^u_i)\right).
\end{equation}
In addition, this process fine-tunes the question correctness tendency estimator, making the tendency score more accurate when jointly trained with the KT model.
It is also worth mentioning that we use this pretraining scheme instead of directly training it because the random initialization of the estimator makes the weights of responses in Equation~\ref{eq:lr-bce} unstable through Equation~\ref{eq:rewrite} and~\ref{eq:d2w}.
This might cause trivial solutions that $w^u_{i}\approx0$ (s.t., $w^u_{i}\geq 0$), and thus lead to a sub-optimal issue.
A pretrained question correctness tendency estimator could provide informative and robust tendency scores at the beginning of the training process.
Besides, the correctness tendency estimator is model-agnostic, allowing new KT models to be directly trained with an existing well-trained estimator, thereby improving efficiency and facilitating the integration of DR4KT into existing KT methods.

\begin{algorithm}[t]
\caption{Training inference procedure of DR4KT.}
\renewcommand{\algorithmicrequire}{ \textbf{Input:}} 
\renewcommand{\algorithmicensure}{ \textbf{Output:}} 
\label{alg:training}
\begin{algorithmic}[1]
\REQUIRE ~~\\ 
    Training set $\mathcal{B}$;
    \\Initialized frequency-aware question correctness tendency estimator $\phi(\cdot|\boldsymbol{\Theta}_\phi)$;
    \\Initialized KT model $\psi(\cdot|\boldsymbol{\Theta}_\psi)$;
    \\Initialized adaptive predictive score fuser  $\kappa(\cdot|\boldsymbol{\Theta}_\kappa)$;
\ENSURE ~~\\
Optimized $\phi(\cdot|\boldsymbol{\Theta}_\phi)$, $\psi(\cdot|\boldsymbol{\Theta}_\psi)$ and $\kappa(\cdot|\boldsymbol{\Theta}_\kappa)$;\\
\STATE Pretrain  $\phi(\cdot|\boldsymbol{\Theta}_\phi)$ via $\mathcal{B}$; (Equation~\ref{eq:pretrain})
\FOR{\textit{number of training epochs}}
\FOR{\textit{sampled mini-batch $B\subset\mathcal{B}$}}
\STATE $\phi(\cdot|\boldsymbol{\Theta}_\phi)$ generates question correctness tendency scores; (Equation~\ref{eq:question} and \ref{eq:diff_score})
\STATE $\psi(\cdot|\boldsymbol{\Theta}_\psi)$ generates KT scores;
\STATE Fetching hidden knowledge state vectors from the KT model;
\STATE Fetching question representations from the question correctness tendency estimator;
\STATE $\kappa(\cdot|\boldsymbol{\Theta}_\kappa)$ fuses predictive scores; (Equation~\ref{eq:disc_gen}, Equation~\ref{eq:fuse} and \ref{eq:final-pre_3})
\STATE Calculate new loss weights; (Equation~\ref{eq:rewrite} and \ref{eq:d2w})
\STATE Update $\boldsymbol{\Theta}_\phi$, $\boldsymbol{\Theta}_\psi$ and $\boldsymbol{\Theta}_\kappa$ by joint training; (Equation~\ref{eq:main_loss})
\ENDFOR
\ENDFOR
\RETURN $\phi(\cdot|\boldsymbol{\Theta}_\phi)$, $\psi(\cdot|\boldsymbol{\Theta}_\psi)$ and $\kappa(\cdot|\boldsymbol{\Theta}_\kappa)$.
\end{algorithmic}
\end{algorithm}

\begin{algorithm}[t]
\caption{Prediction inference procedure of DR4KT.}
\renewcommand{\algorithmicrequire}{ \textbf{Input:}} 
\renewcommand{\algorithmicensure}{ \textbf{Output:}} 
\label{alg:inference}
\begin{algorithmic}[1]
\REQUIRE ~~\\ 
Student $u$'s historical responses $\Hset_t^u$;
\\Target question $q_{t+1}^u$;
\\Frequency-aware question correctness tendency estimator $\phi(\cdot|\boldsymbol{\Theta}_\phi)$;
\\Reweighted KT model $\psi(\cdot|\boldsymbol{\Theta}_\psi)$;
\\Adaptive predictive score fuser $\kappa(\cdot|\boldsymbol{\Theta}_\kappa)$;
\ENSURE ~~\\
Predicted score $\hat{y}_{t+1}^u$ to answer the new question correctly;\\
\STATE $\phi(q_{t+1}^u|\boldsymbol{\Theta}_\phi)$ generates question correctness tendency score $\hat{b}_{t+1}^u$; (Equation~\ref{eq:question} and \ref{eq:diff_score})
\STATE $\psi(\Hset_t^u, q_{t+1}^u|\boldsymbol{\Theta}_\psi)$ generates KT score $\hat{a}_{t+1}^u$;
\STATE Fetching hidden knowledge state vector $\textbf{m}^u_{t+1}$ from the KT model;
\STATE Fetching question representation $\textbf{e}_{t+1}$ from the question correctness tendency estimator;
\STATE $\kappa(\textbf{m}^u_{t+1},\textbf{e}^u_{t+1},\hat{a}_{t+1}^u,\hat{b}_{t+1}^u|\boldsymbol{\Theta}_\kappa)$ fuses predictive score $\hat{y}_{t+1}^u$; (Equation~\ref{eq:disc_gen}, Equation~\ref{eq:fuse} and \ref{eq:final-pre_3})
\RETURN $\hat{y}_{t+1}^u$.
\end{algorithmic}
\end{algorithm}
\subsection{Prediction and Knowledge Tracing}
Once the entire DR4KT framework has been trained, we can use it to predict the probability score of a student answering a target question correctly by using the fused predictive scores from Equation~\ref{eq:fuse}. However, for tracing students' knowledge proficiency, we use only the reweighted KT model, excluding the question correctness tendency estimator and the adaptive predictive score fuser.
This is because the correctness tendency scores obtained from the question correctness tendency estimator are based on the entire student group's responses to each question and do not capture individual students' knowledge states. On the other hand, the reweighted KT model takes into account the response discrimination and focuses on high discriminative responses, which better reflect students' knowledge states.
The way of extracting knowledge proficiency depends on the specific KT model in use.
Take DKT as an example.
Its modeled knowledge proficiency is stored in the $|\mathcal{C}|$-dimension vector that is generated from linearly projecting students' hidden knowledge state vectors, where $|\mathcal{C}|$ is the number of concepts.
Other examples such as AKT, which uses learnable parameters, or EKT, which adds more information in input, are also adapted to our framework.
Details of such extraction can be referred in their original papers~\cite{dkt,akt,ekt}.
The whole training and prediction inference procedures of DR4KT are presented in Algorithm~\ref{alg:training}~and~\ref{alg:inference}.
We omit the batch values' notations in Algorithm~\ref{alg:training} for conciseness.

\subsection{Complexity Analysis}
DR4KT introduces a discrimination rebalancing framework that enhances current KT methods by focusing on highly discriminative responses.
We analyze its complexity to ensure it does not significantly increase in terms of time and space.
Let $t$ be the student's historical sequence length and $d$ the number of model embedding and hidden state dimensions. The number of network layers $L$ is constant and therefore omitted.

The time complexity of DR4KT involves the Hadamard product and fully-connected network in the frequency-aware question correctness tendency estimator with $O(td)$, and the MLP in the adaptive predictive score fusion with $O(td^2)$.
The highest order term gives a total time complexity of $O(td^2)$.
Applied to DKT, SAKT, and AKT, which have original time complexities of $O(td^2), O(t^2d)$ and $O(t^2d)$ respectively, the resulting complexities for DR4KT are $O(td^2)$, $O(t^2d + td^2)$ and $O(t^2d + td^2)$.
Compared to state-of-the-art KT methods like DIMKT and IEKT (both $O(td^2)$), DR4KT's time complexity is similarly acceptable. The space complexity includes problem, concept, and frequency embeddings with $O(td)$, and parameters of the fully-connected network with $O(d^2)$.
The MLP in the adaptive predictive score fusion adds a space complexity of $O(td + d^2)$.
Thus, DR4KT's total space complexity is $O(td + d^2)$, which is in the same order of magnitude as the backbones and state-of-the-art KT methods ($O(td+d^2)$ for DKT, DIMKT and IEKT, and $O(t^2 + td+d^2)$ for SAKT and AKT).
In summary, DR4KT maintains the same computational complexity as most common KT methods while delivering superior knowledge tracing performance.

\section{Experiments}
In this section, we conduct comprehensive experiments on three widely-used datasets to answer the following questions:
\begin{itemize}
	\item[\textbf{\texttt{Q1}:}] How does the proposed discrimination rebalancing framework DR4KT improve knowledge tracing models?
	
	\item[\textbf{\texttt{Q2}:}] What are the contributions of the main components in DR4KT?

        \item[\textbf{\texttt{Q3}:}] Does DR4KT address the discrimination imbalance issue that sacrifices performance on high discriminative responses?
 
\end{itemize}

Additionally, we conduct hyper-parameter analysis and visualize the knowledge states of a real case to provide insights into the knowledge tracing process.

\subsection{Experimental Setup}\label{subsec:exp-setup}

\subsubsection{Datasets}\label{subsec:data} We use three widely-used public datasets with different periods and sizes to validate the efficiency of DR4KT. 

$\bullet$ \textbf{ASSIST09}~\cite{assistments}: This dataset is gathered from an online tutoring system ASSISTments that teaches and accesses students in mathematics. Specifically, we use the \textit{combined dataset} version\footnote{\url{https://sites.google.com/site/assistmentsdata/home/2009-2010-assistment-data/combined-dataset-2009-10}}. 

$\bullet$ \textbf{ASSIST12}~\cite{assistments}: Another dataset from the same platform, but with only one knowledge concept for one question.\footnote{\url{https://sites.google.com/site/assistmentsdata/datasets/2012-13-school-data-with-affect}}.

$\bullet$ \textbf{Eedi}~\cite{eedi}:  This dataset is collected during two school years (2018-2020), with student answers to mathematics questions from Eedi, a free homework and teaching platform for primary and secondary schools in the UK.
We use the \textit{train\_task\_1\_2.csv} as the response dataset in practice. Moreover, the leaf nodes of the provided math concept tree are used as the related knowledge concepts for each question.\footnote{\url{https://eedi.com/projects/neurips-education-challenge}}. 

\begin{table}[t]
\begin{center}
\caption{Statistics of the three datasets after preprocessing.}
\label{tab:dataset}
\begin{tabular}{l|ccc}
\hline
\hline
Dataset                 & ASSIST09 & ASSIST12 & Eedi \\ \hline
collection period       & 2009-2010    & 2012-2013    & 2018-2020 \\
\# of sequences             & 7.3k     & 39.1k    & 206.5k   \\
\# of concepts              & 151      & 265      & 316       \\
\# of responses             & 424.9k   & 2.7m      & 15.8m     \\
\# of questions             & 13.5k     & 53.1k   & 27.6k     \\
avg. passing rate       & 66.5\%      & 68.1\%     & 67.2\% \\ \hline\hline
\end{tabular}
\end{center}
\end{table}

We split each student's historical response sequence into several subsequences with a fixed length of 100. 
We discard any subsequence with fewer than 10 responses. The subsequences with less than 100 responses are padded with zero.
The details of the preprocessed datasets are illustrated in Table~\ref{tab:dataset}.

\subsubsection{Evaluation} We use a five-fold cross validation to evaluate student performance. Furthermore, area under the curve (AUC), accuracy (ACC) and root mean squared error (RMSE) are used as evaluation metrics, which are commonly used in KT tasks. We also apply the early stopping strategy that stops each training process the performance on the validation set does not improve for 10 consecutive epochs.

\subsubsection{Backbones} For comprehensively exhibiting the improvement of the DR4KT framework, we select three commonly-used KT models focusing on different aspects. 

$\bullet$ \textbf{DKT} \cite{dkt} is a milestone method that applies RNN to KT, which captures students' hidden knowledge states and shows improvements over traditional KT methods.

$\bullet$ \textbf{SAKT} \cite{sakt} is an attention-based method employing transformer frameworks in KT. It aims to model the correlations between different responses from the same student, which is in contrast to the RNN architecture.

$\bullet$ \textbf{AKT} \cite{akt} is another transformer-based KT method utilizing the monotonic attention mechanism that shows state-of-the-art performance.
Different from DKT and SAKT, AKT uses question information to enhance performance.

\begin{table}[t]
\begin{center}
\caption{Parameter setting of DR4KT for the three backbones. }
\label{tab:para-dr4kt}
\begin{tabular}{l|cccc|cccc|cccc}
\hline\hline
Dataset   & \multicolumn{4}{c|}{ASSIST09}                   & \multicolumn{4}{c|}{ASSIST12}                   & \multicolumn{4}{c}{Eedi}                        \\ \hline
Parameter & $\tau_1$ & $\tau_2$ & $\lambda_1$ & $\lambda_2$ & $\tau_1$ & $\tau_2$ & $\lambda_1$ & $\lambda_2$ & $\tau_1$ & $\tau_2$ & $\lambda_1$ & $\lambda_2$ \\ \hline
DKT       & 0.5      & 1.0      & 0.5         & 1.0         & 0.5      & 1.0      & 2.0         & 0.2         & 0.2      & 1.0      & 1.0         & 0.5         \\
SAKT      & 0.2      & 1.0      & 2.0         & 1.0         & 0.2      & 2.0      & 1.0         & 1.0         & 0.5      & 1.0      & 1.0         & 1.0         \\
AKT       & 0.2      & 1.0      & 1.0         & 1.0         & 0.2      & 1.0      & 2.0         & 0.5         & 0.2      & 1.0      & 1.0         & 0.5         \\ \hline\hline
\end{tabular}
\end{center}
\end{table}

\subsubsection{Baselines} To demonstrate the superiority of DR4KT in resolving discrimination imbalance and enhancing the overall performance, we compare it with two categories of baselines. The first category consists of the original KT backbones and several alternative frameworks. The second category includes KT methods that explicitly leverage question difficulty or use the thought: answering questions with different difficulty gains different knowledge, which is similar to response discrimination and thus used for comparison.

$\bullet$ \textbf{Original}: The original model of each backbone.

$\bullet$ \textbf{QUES}: A framework baseline that adds question information as input embedding in DKT and SAKT to make fair comparison.

$\bullet$ \textbf{IPW}: The inverse propensity weighting (IPW)~\cite{ipw} framework assigns each loss term an inverse propensity score to eliminate a specified data bias in real environments. IPW is widely used in multiple fields to tackle data imbalance issues. 
In the experiments, we set 10 discrimination levels according to discrimination scores and use their corresponding frequencies as the propensity scores. 

$\bullet$ \textbf{DIFF}: A framework baseline introducing the question difficulty input scheme proposed in the study~\cite{dimkt} to each backbone, thus forming a difficulty-based framework. Likewise, the number of difficulty levels is set to 100.

$\bullet$ \textbf{KT-IDEM}~\cite{kt-idem}: A maching learning based method using Bayesian knowledge tracing incorporating question difficulty parameters.

$\bullet$ \textbf{DIMKT}~\cite{dimkt}: A state-of-the-art method fully utilizing difficulty information. It involves the idea that correctly answering hard questions gains more knowledge, and vice versa.

$\bullet$ \textbf{IEKT}~\cite{iekt}: Another state-of-the-art method baseline that employs individual cognition and acquisition estimation.
It use the idea that students acquiring different knowledge when answering questions with different difficulty.

\begin{table*}[t]
\setlength{\tabcolsep}{2.9pt}
\begin{center}
    
  \caption{Overall performance of all the adopted backbones, frameworks and baselines. The best results for each metric, across all framework baselines for each backbone, are in bold. The second-best results for each backbone are in italic. The best results among all baselines are underlined. The three attributes ``Q'', ``D'' and ``RD'' indicate whether the method uses question information, difficulty information and response discrimination. The percentages in the middle show the improvement over the best baseline for each backbone. The percentages in the last row show the improvement of the best DR4KT result over the best baseline overall.}
\label{tab:main_exp}
\begin{tabular}{l|ccc|ccc|ccc|ccc}
\hline
\hline
\multirow{2}{*}{Model} & \multicolumn{3}{c|}{Attribute}                                      & \multicolumn{3}{c|}{ASSIST09}      & \multicolumn{3}{c|}{ASSIST12}      & \multicolumn{3}{c}{Eedi}          \\ \cline{2-13} 
                       & Q                & D                & RD        & AUC$\uparrow$             & ACC$\uparrow$ & RMSE$\downarrow$            & AUC$\uparrow$             & ACC$\uparrow$ & RMSE$\downarrow$            & AUC$\uparrow$             & ACC$\uparrow$    & RMSE$\downarrow$         \\ \hline
\textbf{DKT}                    &                      &                      &                       & 0.7708          & 0.7248    & 0.4276      & 0.7298          & 0.7345     & 0.4245     & 0.7433          & 0.7049   & 0.4381       \\
\,\,+QUES               & $\checkmark$            &                      &                       & \textit{0.7743}          & \textit{0.7264}        & 0.4275    & \textit{0.7337}        & 0.7360       & 0.4233   & \textit{0.7518}          & \textit{0.7094}     & 0.4349     \\
\,\,+IPW                &                      &                      & $\checkmark$             & 0.7289          & 0.6844      & 0.4529    & 0.6700          & 0.6942      & 0.4637    & 0.7176          & 0.6699     & 0.4592     \\
\,\,+DIFF               & $\checkmark$            & $\checkmark$            &                       & 0.7736          & 0.7262    & 0.4258      & 0.7327          & \textit{0.7362}      & 0.4235    & 0.7498          & 0.7086    & 0.4367      \\ \hline
\,\,+DR4KT              & $\checkmark$            & $\checkmark$            & $\checkmark$             & \textbf{0.7891}*$^\mathrm{b}$ & \textbf{0.7409}* & \textbf{0.4209}* & \textbf{0.7707}* & \textbf{0.7534} & \textbf{0.4117}* & \textbf{0.7842}* & \textbf{0.7294}* & \textbf{0.4222}*\\
improv.                & \multicolumn{1}{l}{} & \multicolumn{1}{l}{} & \multicolumn{1}{l|}{} & \textit{+1.92\%} & \textit{+1.99\%} & \textit{+1.15\%}& \textit{+5.05\%} & \textit{+2.33\%} & \textit{+2.74\%}& \textit{+4.30\%}* & \textit{+2.82\%} & \textit{+2.92\%} \\ \hline\hline
\textbf{SAKT}                   &                      &                      &                       & 0.7575          & 0.7156    & 0.4327      & 0.7218          & 0.7314      & 0.4271    & 0.7452          & 0.7061     & 0.4376     \\
\,\,+QUES              & $\checkmark$            &                      &                       & 0.7737          & 0.7265     & 0.4287     & 0.7433          & 0.7381     & 0.4234     & 0.7754          & 0.7238     & 0.4329     \\
\,\,+IPW               &                      &                      & $\checkmark$             & 0.7103          & 0.6755      & 0.4596    & 0.6661          & 0.6831      & 0.4636    & 0.7207          & 0.6710      & 0.4579    \\
\,\,+DIFF              & $\checkmark$            & $\checkmark$            &                       & \textit{0.7747}          & \textit{0.7282}  & 0.4269        & \textit{0.7582}          & \textit{0.7469}    & 0.4218      & \textit{0.7774}          & \textit{0.7250} & 0.4315         \\ \hline
\,\,+DR4KT             & $\checkmark$            & $\checkmark$            & $\checkmark$             & \textbf{0.7860}* & \textbf{0.7349}* & \textbf{0.4234}* & \textbf{0.7662}* & \textbf{0.7507}* & \textbf{0.4179}* & \textbf{0.7845}* & \textbf{0.7298}* & \textbf{0.4209}* \\
improv.                & \multicolumn{1}{l}{} & \multicolumn{1}{l}{} & \multicolumn{1}{l|}{} & \textit{+1.46\%} & \textit{+0.91\%} & \textit{+0.82\%} & \textit{+1.05\%} & \textit{+0.51\%} & \textit{+0.92\%} & \textit{+0.91\%} & \textit{+0.66\%} & \textit{+2.46\%} \\ \hline\hline
\textbf{AKT}                    & $\checkmark$            & $\circ^{\mathrm{a}}$           &                       & 0.7840          & 0.7344 & 0.4242         & 0.7626          & 0.7490   & 0.4149       & 0.7882          & 0.7325   & 0.4230      \\
\,\,+IPW                & $\checkmark$            & $\circ$           & $\checkmark$             & 0.7430          & 0.7013      & 0.4495    & 0.6938          & 0.7157       & 0.4747   & 0.7405          & 0.6796      & 0.4548    \\
\,\,+DIFF               & $\checkmark$            & $\checkmark$            &                       & \underline{\textit{0.7856}}          & \textit{0.7351}      & 0.4240    & \textit{0.7637}          & \textit{0.7497} & 0.4140         & \underline{\textit{0.7893}}          & \textit{0.7329}      & \underline{0.4210}    \\ \hline
\,\,+DR4KT              & $\checkmark$            & $\checkmark$            & $\checkmark$             & \textbf{0.7919}**${^\mathrm{c}}$ & \textbf{0.7425}** & \textbf{0.4217}** & \textbf{0.7714}** & \textbf{0.7561}** & \textbf{0.4107}* & \textbf{0.7946}** & \textbf{0.7367}** & \textbf{0.4196}**\\
improv.                & \multicolumn{1}{l}{} & \multicolumn{1}{l}{} & \multicolumn{1}{l|}{} & \textit{+0.80\%} & \textit{+1.01\%} & \textit{+0.54\%} & \textit{+1.01\%} & \textit{+0.85\%} & \textit{+0.80\%} & \textit{+0.67\%} & \textit{+0.52\%} & \textit{+0.33\%} \\ \hline\hline
KT-IDEM                & $\checkmark$            & $\checkmark$            &                       & 0.7292          & 0.6951      & 0.4494    & 0.7102          & 0.7239       & 0.4323   & 0.7101          & 0.6840       & 0.4485   \\
DIMKT                  & $\checkmark$            & $\checkmark$            & $\diamond^\mathrm{d}$             & 0.7815          & 0.7351       & 0.4254   & \underline{0.7688}          & \underline{0.7531}       & \underline{0.4120}   & 0.7888          & \underline{0.7330}   & 0.4211       \\
IEKT                   & $\checkmark$            &                      & $\diamond$             & 0.7835          & \underline{0.7353}    & \underline{0.4238}      & 0.7671          & 0.7517     & 0.4239     & 0.7835          & 0.7286     & 0.4314     \\ \hline
improv.                &                      &                      &                       & \textit{+0.80\%} & \textit{+0.98\%} & \textit{+0.50\%} & \textit{+0.34\%} & \textit{+0.40\%} & \textit{+0.32\%} & \textit{+0.67\%} & \textit{+0.50\%} & \textit{+0.33\%} \\ \hline\hline
\multicolumn{13}{l}{}\\
\multicolumn{13}{l}{$^{\mathrm{a}}$ $\circ$ means AKT does not explicitly use the difficulty information.}\\
\multicolumn{13}{l}{$^{\mathrm{b}}$ * indicates statistical significance over the best baseline of the corresponding backbone by T-test with $p\leq0.05$.}\\
\multicolumn{13}{l}{$^{\mathrm{c}}$ ** indicates statistical significance over the best result of all the baselines.} \\
\multicolumn{13}{l}{$^{\mathrm{d}}$ $\diamond$ means the similar thought to response discrimination.} \\
\end{tabular}
\end{center}
\end{table*}


\subsubsection{Implementation details} 
All the experiments are conducted on a Linux server with GPUs of GeForce GTX 2080Ti under the deep learning framework Pytorch. For all baselines with open source code, we directly duplicate them and only modify necessary function arguments to fit our code frameworks. For those without open-source code, we strictly reproduce their models according to the original papers.
We set the embedding and hidden dimension size of each module to 128 for efficiency and fair comparison.
In the training stage, we use the Adam optimizer~\cite{adam} and set the batch size to 128.
For the backbones, we mainly tune the universal parameters such as the learning rate, dropout ratio, and $l_2$ regularization values.
The remaining model-specific hyper-parameters are strictly set according to the original papers.
All the baselines and backbones are tuned to their best performance.
The four hyper-parameters of DR4KT are all selected from \{0.1,0.2,0.5,1.0,2.0,5.0\}, and the final settings are presented in Table~\ref{tab:para-dr4kt}.
In addition, the learning rates to pretrain the frequency-aware correctness tendency estimator of the three datasets are set to \{0.02,0.01,0.005\}. The numbers of the frequency slots are set to \{40,20,20\}.
The numbers of embedding and hidden dimensions in the modules of DR4KT are also set to 128, including both the frequency-aware question correctness tendency estimator and the adaptive predictive score fusion.

\begin{table}[t]

\centering
\caption{The result of ablation study of DR4KT applied to three backbones.}
\label{tab:exp-ablation}
\begin{tabular}{l|ccc|ccc|ccc}
\hline\hline
Dataset    & \multicolumn{3}{c|}{ASSIST09}      & \multicolumn{3}{c|}{ASSIST12}      & \multicolumn{3}{c}{Eedi}          \\ \hline
Model      & AUC$\uparrow$             & ACC$\uparrow$      & RMSE$\downarrow$       & AUC$\uparrow$             & ACC$\uparrow$     & RMSE$\downarrow$        & AUC$\uparrow$             & ACC$\uparrow$    & RMSE$\downarrow$         \\ \hline
DKT+DR4KT  & \textbf{0.7891} & \textbf{0.7409} & \textbf{0.4209} & \textbf{0.7707} & \textbf{0.7534} & 0.4117 & \textbf{0.7842} & \textbf{0.7294} & 0.4222 \\
-TE+FQ   & 0.7870          & 0.7379     & 0.4233     & 0.7661          & 0.7502          & 0.4178 & 0.7826          & 0.7280        & 0.4313  \\
-LR        & 0.7832          & 0.7372      & 0.4239    & 0.7655          & 0.7504      & 0.4156    & 0.7789          & 0.7262      & 0.4231    \\
-LR\_LOSS        & 0.7855          & 0.7380  & 0.4214        & 0.7670          & 0.7518   & 0.4133      & 0.7814          & 0.7266    & 0.4237      \\
-DC\_LOSS        & 0.7872          & 0.7388  & 0.4229        & 0.7691          & 0.7521    & 0.4130      & 0.7790          & 0.7268   & 0.4249       \\ \hline
SAKT+DR4KT & \textbf{0.7860} & \textbf{0.7349} & \textbf{0.4234} & \textbf{0.7662} & \textbf{0.7507} & \textbf{0.4179} & \textbf{0.7845} & \textbf{0.7298} & \textbf{0.4209} \\
-TE+FQ   & 0.7838          & 0.7323      & 0.4275    & 0.7630          & 0.7485    & 0.4205      & 0.7829          & 0.7278      & 0.4318    \\
-LR        & 0.7778          & 0.7304    & 0.4254      & 0.7600          & 0.7475    & 0.4233      & 0.7783          & 0.7265    & 0.4259      \\
-LR\_LOSS        & 0.7785          & 0.7315 & 0.4250         & 0.7605          & 0.7482   & 0.4227       & 0.7804          & 0.7253    & 0.4257      \\
-DC\_LOSS        & 0.7843          & 0.7340  & 0.4241        & 0.7646          & 0.7498    & 0.4199      & 0.7820          & 0.7274    & 0.4229      \\ \hline
AKT+DR4KT  & \textbf{0.7919} & \textbf{0.7425} & \textbf{0.4217} & \textbf{0.7714} & \textbf{0.7561} & \textbf{0.4107} & \textbf{0.7946} & \textbf{0.7367} & \textbf{0.4196} \\
-TE+FQ   & 0.7864          & 0.7386     & 0.4243     & 0.7673          & 0.7541       & 0.4159   & 0.7903          & 0.7349        & 0.4250  \\
-LR        & 0.7847          & 0.7371   & 0.4317       & 0.7650          & 0.7533      & 0.4135    & 0.7872          & 0.7328       & 0.4235   \\
-LR\_LOSS        & 0.7825          & 0.7350  & 0.4238        & 0.7663          & 0.7538   & 0.4128       & 0.7901          & 0.7338    & 0.4217      \\
-DC\_LOSS         & 0.7891          & 0.7392    & 0.4261      & 0.7698          & 0.7548    & 0.4121      & 0.7912          & 0.7341     & 0.4215     \\ \hline\hline
\end{tabular}

\end{table}

\subsection{Overall Performance (\textbf{\texttt{Q1}})}
The experimental results in Table~\ref{tab:main_exp} show the superiority of the proposed DR4KT framework compared to other baseline methods and frameworks. DR4KT consistently outperforms all the other baselines and frameworks on all three datasets, with performance improvements ranging from 0.32\% to 5.05\% of the metrics. This significant improvement highlights the effectiveness of DR4KT in enhancing knowledge tracing models.
Among the different backbone models, AKT achieves relatively higher performance, while DKT and SAKT are inferior. This performance trend remains consistent even after applying DR4KT to these backbones. DR4KT provides the highest performance enhancement for SAKT on the ASSIST12 dataset, achieving a 5.60\% increase in AUC. The lowest improvement is observed on Eedi with the AKT backbone, showing a 0.57\% increase in ACC.
When comparing DR4KT with other frameworks, the IPW framework degrades the performance compared to the original backbones. This is because IPW prioritizes higher discriminative responses, sacrificing the performance on lower discriminative responses. On the other hand, the DIFF framework, which only leverages difficulty information, shows improvement on all backbones compared to their original versions but is still inferior to DR4KT. This suggests that DR4KT, which considers both difficulty and response discrimination, is more effective in enhancing knowledge tracing models.
Additionally, DKT and SAKT with additional question embedding show improvement, indicating the value of incorporating question information.
Overall, the results highlight the generality and effectiveness of the proposed DR4KT framework in addressing the discrimination imbalance issue in KT. DR4KT provides a more comprehensive and balanced approach by considering both difficulty and response discrimination, leading to improved performance across different backbone models and datasets.

\subsection{Ablation Study (\textbf{\texttt{Q2}})}
To assess the impact of DR4KT's individual components on overall performance, we conducted an ablation study, detailed in Table~\ref{tab:exp-ablation}. The suffix ``-LR'' denotes the exclusion of loss reweighting for discriminative responses in the original KT backbones, meaning that all weights are set to 1.
Additionally, ``-LR\_LOSS'' signifies the removal of the total loss reweighting objective function in joint training. Similarly, ``-DC\_LOSS'' involves omitting the loss aligning calculated response discrimination scores with generated discrimination scores in the adaptive predictive score fuser.
Furthermore, to showcase DR4KT's ability to address the sparsity issue in questions, we replaced generated question correctness tendency scores with question passing rates. We also excluded the question representation part in the concatenation of the adaptive predictive score fusion. This ablation is denoted as ``-TE+FQ''.
As observed, ``-LR\_LOSS'' and ``-DC\_LOSS'' resulted in a performance decline compared to the full DR4KT framework, indicating that both auxiliary tasks - reweighting KT models and discrimination score alignment - contribute positively to overall performance. Notably, ``-LR\_LOSS'' led to a more substantial decrease, emphasizing the importance of the loss reweighting task for DR4KT.
In comparison to other ablation trails, ``-DC\_LOSS'' did not exhibit significant degradation. This could be attributed to the fact that a portion of the response discrimination scores corresponds to ground-truth response correctness, which is challenging for models to learn.
Additionally, ``-LR'' also caused a performance decrease but was inferior to ``-LR\_LOSS'' in most cases. This suggests that having only an auxiliary task to train the KT model with DR4KT is counterproductive, highlighting the necessity of a loss reweighting scheme.
Moreover, the performance of ``-TE+FQ'' also declined, confirming that the pretrained correctness tendency estimator effectively addresses the question sparsity issue. It provides more accurate and noiseless correctness tendency scores, beneficial for adaptive predictive score fusion.
Notably, the degradation in performance is smaller on the Eedi dataset. This could be attributed to Eedi's large dataset size, which mitigates the question sparsity issue, making direct approximation less inexact.
Overall, the ablation study confirms the effectiveness and significance of each component in the DR4KT framework.

\begin{table}[!t]
\centering
\caption{Balance analysis of DR4KT applied to four backbones on the two ASSIST datasets.}
\label{tab:exp-balance}
\begin{tabular}{lcccccc}
\hline\hline
\multicolumn{1}{l|}{ASSIST09}         & \multicolumn{6}{c}{Discrimination Level of Responses}                                                                          \\ \hline
\multicolumn{1}{l|}{Model}            & \multicolumn{1}{c|}{Overall}         & {[}0, 0.2)      & {[}0.2, 0.4)    & {[}0.4, 0.6)    & {[}0.6, 0.8)    & {[}0.8, 1.0{]}  \\ \hline
\multicolumn{1}{l|}{DKT}              & \multicolumn{1}{c|}{0.7248}          & 0.9247          & 0.8197          & 0.6494          & 0.4477          & 0.2959          \\
\multicolumn{1}{l|}{DKT+TE}              & \multicolumn{1}{c|}{0.6335}          & 0.9793	& 0.7832	& 0.4860	& 0.2062	& 0.0231
         \\
\multicolumn{1}{l|}{DKT+LR}           & \multicolumn{1}{c|}{0.6907}          & 0.7738	&0.7482	&0.6543	&0.5452	&0.5275
          \\
\multicolumn{1}{l|}{\textbf{DKT+DR}}  & \multicolumn{1}{c|}{\textbf{0.7409}} & \textbf{0.9694} & \textbf{0.8599} & \textbf{0.6503} & \textbf{0.4016} & \textbf{0.2378} \\ \hline
\multicolumn{1}{l|}{SAKT}             & \multicolumn{1}{c|}{0.7156}          & 0.9172          & 0.8086          & 0.6378          & 0.4416          & 0.2895          \\
\multicolumn{1}{l|}{SAKT+TE}              & \multicolumn{1}{c|}{0.6305}          & 0.9750	& 0.7744	&0.4946	&0.1989&	0.0257
         \\
\multicolumn{1}{l|}{SAKT+LR}          & \multicolumn{1}{c|}{0.6997}          & 0.8378	&0.7554	&0.6474	&0.5260	&0.4549
         \\
\multicolumn{1}{l|}{\textbf{SAKT+DR}} & \multicolumn{1}{c|}{\textbf{0.7349}} & \textbf{0.9698} & \textbf{0.8723} & \textbf{0.6448} & \textbf{0.3754} & \textbf{0.2017} \\ \hline
\multicolumn{1}{l|}{AKT}              & \multicolumn{1}{c|}{0.7334}          & 0.9715          & 0.8735          & 0.6465          & 0.3579          & 0.1736          \\
\multicolumn{1}{l|}{AKT+TE}              & \multicolumn{1}{c|}{0.6749}          & 0.9999	&0.9722	&0.4977	&0.0193&	0.0001          \\
\multicolumn{1}{l|}{AKT+LR}           & \multicolumn{1}{c|}{0.7059}          & 0.8481	&0.7640	&0.6516	&0.5262	&0.4532          \\
\multicolumn{1}{l|}{\textbf{AKT+DR}}  & \multicolumn{1}{c|}{\textbf{0.7425}} & \textbf{0.9730} & \textbf{0.8977} & \textbf{0.6450} & \textbf{0.3654} & \textbf{0.1921} \\ \hline
\multicolumn{1}{l|}{DIMKT}  & \multicolumn{1}{c|}{0.7351} & 0.9864 & 0.9040 & 0.6449& 0.2988 & 0.1079 \\
\multicolumn{1}{l|}{IEKT}  & \multicolumn{1}{c|}{0.7353} & 0.9901 & 0.9147 & 0.6421& 0.2797 & 0.0876\\\hline\hline
                                      &                                      &                 &                 &                 &                 &                 \\ \hline\hline
\multicolumn{1}{l|}{ASSIST12}         & \multicolumn{6}{c}{Discrimination Level of Responses}                                                                          \\ \hline
\multicolumn{1}{l|}{Model}            & \multicolumn{1}{c|}{Overall}         & {[}0, 0.2)      & {[}0.2, 0.4)    & {[}0.4, 0.6)    & {[}0.6, 0.8)    & {[}0.8, 1.0{]}  \\ \hline
\multicolumn{1}{l|}{DKT}              & \multicolumn{1}{c|}{0.7347}          & 0.9414          & 0.8739          & 0.6142          & 0.3103          & 0.2051          \\
\multicolumn{1}{l|}{DKT+TE}              & \multicolumn{1}{c|}{0.7041}          & 0.9978	&0.9295	&0.4900	&0.0775	&0.0020
          \\
\multicolumn{1}{l|}{DKT+LR}           & \multicolumn{1}{c|}{0.6524}          & 0.7446	&0.6292	&0.6139	&0.6029	&0.5471
          \\
\multicolumn{1}{l|}{\textbf{DKT+DR}}  & \multicolumn{1}{c|}{\textbf{0.7534}} & \textbf{0.9821} & \textbf{0.9218} & \textbf{0.6136} & \textbf{0.2631} & \textbf{0.1381} \\ \hline
\multicolumn{1}{l|}{SAKT}             & \multicolumn{1}{c|}{0.7314}          & 0.9412          & 0.8750          & 0.6083          & 0.2986          & 0.1923          \\
\multicolumn{1}{l|}{SAKT+TE}              & \multicolumn{1}{c|}{0.7054}          & 0.9987	&0.9393	&0.4879	&0.0651&	0.0010
          \\
\multicolumn{1}{l|}{SAKT+LR}          & \multicolumn{1}{c|}{0.6789}          & 0.8019	&0.7097	&0.6084	&0.5209	&0.4492
          \\
\multicolumn{1}{l|}{\textbf{SAKT+DR}} & \multicolumn{1}{c|}{\textbf{0.7507}} & \textbf{0.9793} & \textbf{0.9230} & \textbf{0.6110} & \textbf{0.2576} & \textbf{0.1275} \\ \hline
\multicolumn{1}{l|}{AKT}              & \multicolumn{1}{c|}{0.7490}          & 0.9924          & 0.9324          & 0.6209          & 0.2107          & 0.0556          \\
\multicolumn{1}{l|}{AKT+TE}              & \multicolumn{1}{c|}{0.7082}          &0.9971	&0.9461	&0.4973	&0.0588	&0.0026
          \\
\multicolumn{1}{l|}{AKT+LR}           & \multicolumn{1}{c|}{0.6600}          & 0.7504	&0.6991	&0.5788	&0.5338	&0.5306
         \\
\multicolumn{1}{l|}{\textbf{AKT+DR}}  & \multicolumn{1}{c|}{\textbf{0.7561}} & \textbf{0.9893} & \textbf{0.9298} & \textbf{0.6244} & \textbf{0.2443} & \textbf{0.1017} \\ \hline
\multicolumn{1}{l|}{DIMKT}  & \multicolumn{1}{c|}{0.7531} & 0.9936 & 0.9246 & 0.6328& 0.2359 & 0.0679\\
\multicolumn{1}{l|}{IEKT}  & \multicolumn{1}{c|}{0.7517} & 0.9945 & 0.9374 & 0.6250& 0.2097 & 0.0598\\ \hline\hline
\end{tabular}
\end{table}

\begin{figure}[!t]
  \centering
  \includegraphics[width=\linewidth]{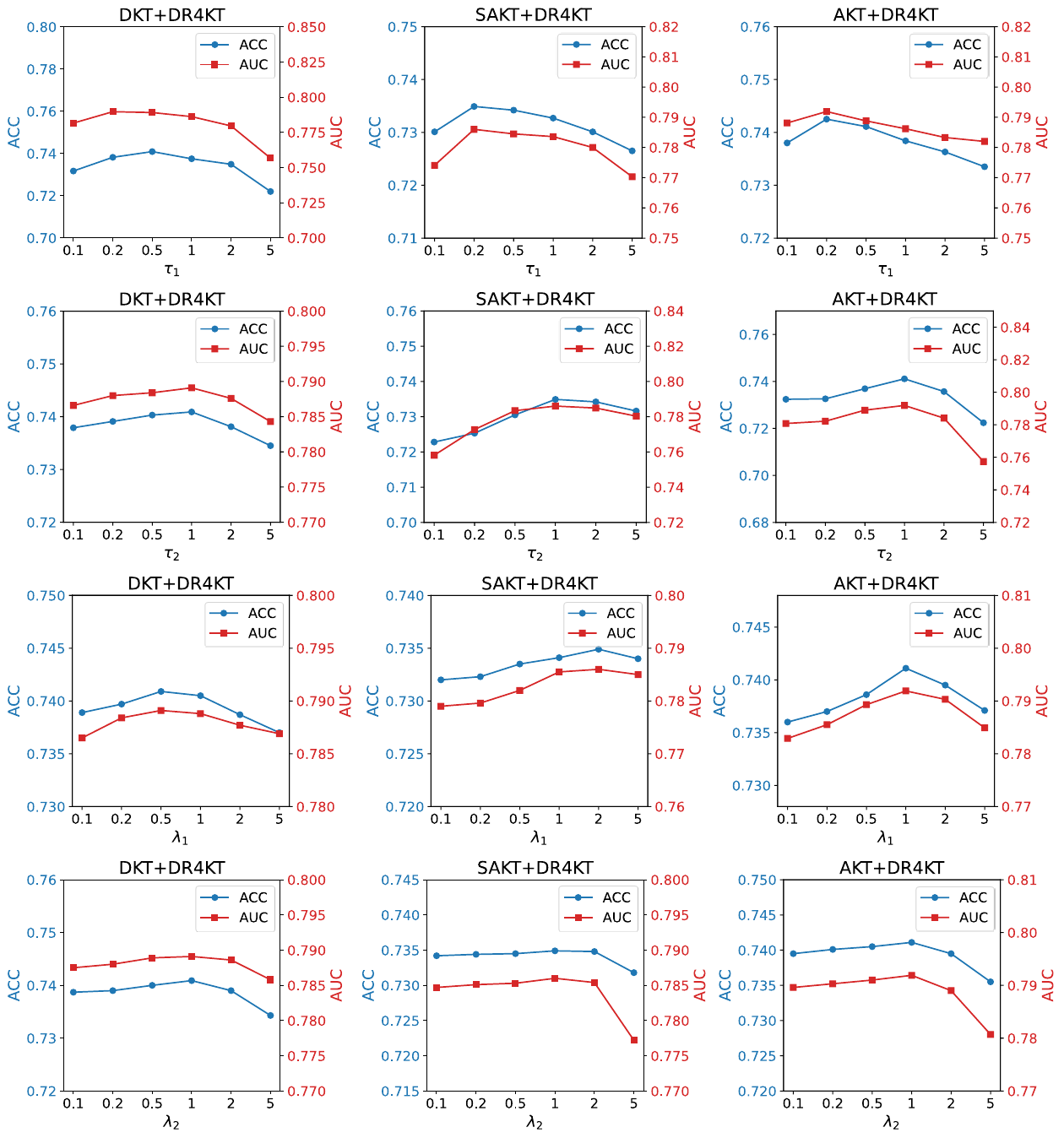}
  \caption{Hyper-parameter analysis of DR4KT on ASSIST09.
  }
 \label{fig:temp}
\end{figure}

\subsection{Balance Analysis (\textbf{\texttt{Q3}})}
In Table~\ref{tab:exp-balance}, we present the detailed performance in terms of accuracy (ACC) on different discriminative levels to validate the effectiveness of rebalancing discriminative responses using our DR4KT framework.
We compare our method with several other approaches. ``KT+TE'' represents the prediction based solely on the generated correctness tendency scores from the fine-tuned frequency-aware question correctness tendency estimator, which already achieves good accuracy, indicating the presence of an imbalance issue in the KT datasets.
``KT+LR'' indicates the prediction using KT backbones trained with our loss reweighting scheme, which alleviates the performance decline on higher discriminative responses, especially compared with the two baselines, DIMKT and IEKT. On the other hand, ``KT-DR'' represents the backbones using our integrated DR4KT framework.
The results show that the degeneration of reweighted models' prediction on low discriminative responses is compensated in the adaptive predictive score fusion by the question correctness tendency estimator, resulting in extremely high accuracy on low discrimination responses (close to 1).
As a result, the overall performance is significantly enhanced, demonstrating the effectiveness of our DR4KT framework in addressing the discrimination imbalance issue. Similar results are observed on the Eedi dataset.

\subsection{Hyper-parameter Analysis}
In the DR4KT model, four hyper-parameters play distinct roles, each contributing to the model's performance. This section presents a thorough hyper-parameter analysis, investigating their impacts by varying values from 0.1 to 5 with increments of \{0.1, 0.2, 0.5, 1, 2, 5\}. The results on ASSIST09 with three different backbones are visualized in Figure~\ref{fig:temp}.
The first parameter, $\tau_1$, governs the intensity of loss reweighting. As depicted in the first row, the performance of DR4KT across the three backbones peaks around 0.2. This highlights that an optimal intensity of loss reweighting is crucial for achieving favorable outcomes.
The second row demonstrates the effectiveness of $\tau_2$," which dictates the adaptive fusion of output scores from the reweighted KT model and the question correctness tendency estimator. Results indicate that when $\tau_2$ is approximately 1.0, signifying an equal combination of scores based on response discrimination, the model yields optimal performance.
In contrast, the third row illustrates varied performance patterns concerning the three backbones, with peak results at 0.5, 2.0, and 1.0, respectively. This underscores the need for different constraints in the loss reweighting scheme for diverse backbones.
Lastly, the parameter governing discrimination scores alignment also proves beneficial for DR4KT. As depicted in the last row, results show a gradual increase initially but decline sharply when the value surpasses 2.0. This indicates that a too high value of $\lambda_2$ might lead to the dominance of this alignment, adversely affecting the learning of the original real KT task.



\begin{figure*}[t]
  \centering
\includegraphics[width=\linewidth]{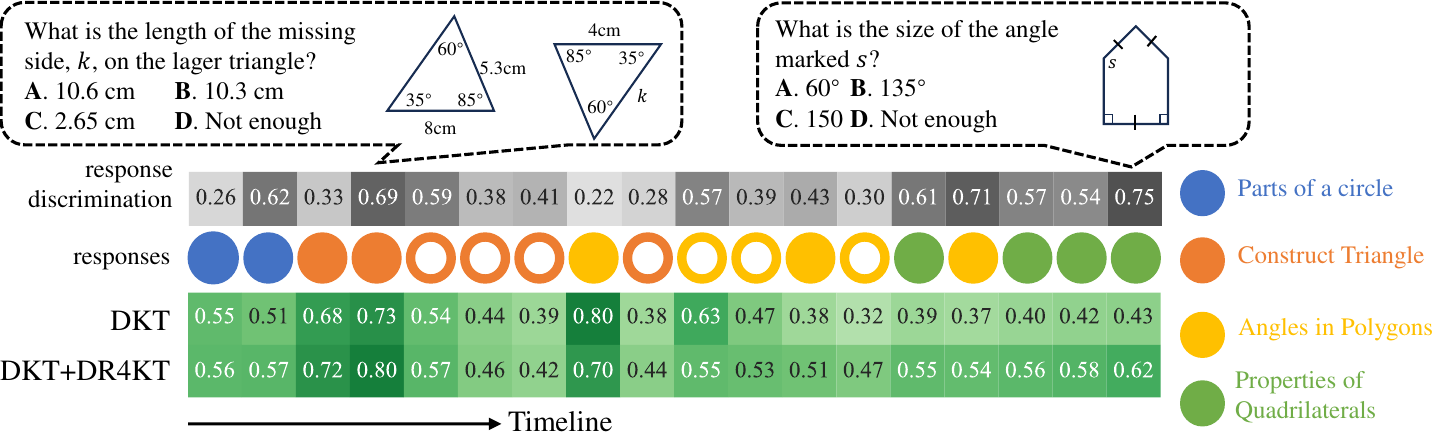}
  \caption{An example of DKT with and without DR4KT tracing student knowledge mastery. We choose a student's response sequence from the Eedi dataset. The gray squares indicate the calculated discrimination scores of each response. Each green square denotes the student's updated knowledge mastery of the corresponding knowledge concept after completing the response. Solid circles represent correct responses and hollow ones represent incorrect responses.}
 \label{fig:kt}
\end{figure*}

\subsection{Knowledge Mastery Visualization}

To validate the effectiveness of DR4KT in focusing on higher discriminative responses, we compare an Eedi student's knowledge states modeled by DKT and its DR4KT version. For better illustration, we use the \textit{train\_task\_3\_4.csv} file, which provides question descriptions.
As shown in Figure~\ref{fig:kt}, the response sequence of the student contains a proportion of high discriminative responses. For example, the last few correct but high discriminative responses suggest that the student has correctly answered these difficulty questions, which implies a high proficiency in the concept \textit{Properties of Quadrilaterals}.
However, this high mastery level is not captured by the original DKT model (with scores ranging from 0.40 to 0.43), but is well traced by the DR4KT version (with scores ranging from 0.54 to 0.62).
Moreover, the traced knowledge mastery of the DR4KT version increases or decreases more when the student has made highly discriminative correct or incorrect responses, which aligns with our intuition (e.g., the fourth response to a difficult question).
This suggests that our DR4KT provides better knowledge tracing on high discriminative responses, which is crucial for KT in discriminating students with different knowledge mastery levels.

\section{Conclusion}

In this paper, we underscore the crucial role of discriminative responses in KT and brings attention to the prevalent issue of response discrimination imbalance. Through meticulous data analysis, we unveil that current KT methods tend to prioritize lower discriminative responses, creating a misleading spike in prediction accuracy that undermines the true knowledge tracing capability.
To tackle this challenge, we introduce DR4KT, an innovative discrimination rebalancing framework designed to reweight responses in KT scenarios. Rigorous experiments showcase that DR4KT remarkably enhances the performance of various KT models. Importantly, it effectively mitigates the performance decline observed in high discriminative responses, leading to a more robust and accurate knowledge tracing process.
In summary, our research highlights the significance of addressing discriminative responses in KT and introduces DR4KT as a practical and potent solution to elevate its overall performance. This work contributes valuable insights to the KT domain and provides a tangible approach for improving the effectiveness of knowledge tracing systems.

\begin{acks}
 The authors would like to thank the valuable comments of editors and reviewers.
\end{acks}


\bibliographystyle{ACM-Reference-Format}
\bibliography{sample-base}


\begin{thebibliography}{52}


\ifx \showCODEN    \undefined \def \showCODEN     #1{\unskip}     \fi
\ifx \showDOI      \undefined \def \showDOI       #1{#1}\fi
\ifx \showISBNx    \undefined \def \showISBNx     #1{\unskip}     \fi
\ifx \showISBNxiii \undefined \def \showISBNxiii  #1{\unskip}     \fi
\ifx \showISSN     \undefined \def \showISSN      #1{\unskip}     \fi
\ifx \showLCCN     \undefined \def \showLCCN      #1{\unskip}     \fi
\ifx \shownote     \undefined \def \shownote      #1{#1}          \fi
\ifx \showarticletitle \undefined \def \showarticletitle #1{#1}   \fi
\ifx \showURL      \undefined \def \showURL       {\relax}        \fi
\providecommand\bibfield[2]{#2}
\providecommand\bibinfo[2]{#2}
\providecommand\natexlab[1]{#1}
\providecommand\showeprint[2][]{arXiv:#2}

\bibitem[Belli et~al\mbox{.}(1997)]%
        {diff-rate}
\bibfield{author}{\bibinfo{person}{Robert~F Belli}, \bibinfo{person}{Daniel~H Hill}, {and} \bibinfo{person}{A~Regula Herzog}.} \bibinfo{year}{1997}\natexlab{}.
\newblock \showarticletitle{Question difficulty and respondents' cognitive ability: The effect on data quality}.
\newblock \bibinfo{journal}{\emph{JOURNAL OF OFFICIAL STATISTICS-STOCKHOLM-}}  \bibinfo{volume}{13} (\bibinfo{year}{1997}), \bibinfo{pages}{181--199}.
\newblock


\bibitem[Chawla et~al\mbox{.}(2002)]%
        {smote}
\bibfield{author}{\bibinfo{person}{Nitesh~V Chawla}, \bibinfo{person}{Kevin~W Bowyer}, \bibinfo{person}{Lawrence~O Hall}, {and} \bibinfo{person}{W~Philip Kegelmeyer}.} \bibinfo{year}{2002}\natexlab{}.
\newblock \showarticletitle{SMOTE: synthetic minority over-sampling technique}.
\newblock \bibinfo{journal}{\emph{Journal of artificial intelligence research}}  \bibinfo{volume}{16} (\bibinfo{year}{2002}), \bibinfo{pages}{321--357}.
\newblock


\bibitem[Chen et~al\mbox{.}(2018)]%
        {pdkt}
\bibfield{author}{\bibinfo{person}{Penghe Chen}, \bibinfo{person}{Yu Lu}, \bibinfo{person}{Vincent~W Zheng}, {and} \bibinfo{person}{Yang Pian}.} \bibinfo{year}{2018}\natexlab{}.
\newblock \showarticletitle{Prerequisite-driven deep knowledge tracing}. In \bibinfo{booktitle}{\emph{2018 IEEE International Conference on Data Mining (ICDM)}}. \bibinfo{pages}{39--48}.
\newblock


\bibitem[Corbett and Anderson(1994)]%
        {kt}
\bibfield{author}{\bibinfo{person}{Albert~T Corbett} {and} \bibinfo{person}{John~R Anderson}.} \bibinfo{year}{1994}\natexlab{}.
\newblock \showarticletitle{Knowledge tracing: Modeling the acquisition of procedural knowledge}.
\newblock \bibinfo{journal}{\emph{User modeling and user-adapted interaction}}  \bibinfo{volume}{4} (\bibinfo{year}{1994}), \bibinfo{pages}{253--278}.
\newblock


\bibitem[Covington et~al\mbox{.}(2016)]%
        {youtube}
\bibfield{author}{\bibinfo{person}{Paul Covington}, \bibinfo{person}{Jay Adams}, {and} \bibinfo{person}{Emre Sargin}.} \bibinfo{year}{2016}\natexlab{}.
\newblock \showarticletitle{Deep neural networks for youtube recommendations}. In \bibinfo{booktitle}{\emph{Proceedings of the 10th ACM conference on recommender systems}}. \bibinfo{pages}{191--198}.
\newblock


\bibitem[Cui et~al\mbox{.}(2023)]%
        {mrtkt}
\bibfield{author}{\bibinfo{person}{Jiajun Cui}, \bibinfo{person}{Zeyuan Chen}, \bibinfo{person}{Aimin Zhou}, \bibinfo{person}{Jianyong Wang}, {and} \bibinfo{person}{Wei Zhang}.} \bibinfo{year}{2023}\natexlab{}.
\newblock \showarticletitle{Fine-Grained Interaction Modeling with Multi-Relational Transformer for Knowledge Tracing}.
\newblock \bibinfo{journal}{\emph{ACM Transactions on Information Systems}} \bibinfo{volume}{41}, \bibinfo{number}{4} (\bibinfo{year}{2023}), \bibinfo{pages}{1--26}.
\newblock


\bibitem[Cui et~al\mbox{.}(2024a)]%
        {cui2024}
\bibfield{author}{\bibinfo{person}{Jiajun Cui}, \bibinfo{person}{Hong Qian}, \bibinfo{person}{Bo Jiang}, {and} \bibinfo{person}{Wei Zhang}.} \bibinfo{year}{2024}\natexlab{a}.
\newblock \showarticletitle{Leveraging Pedagogical Theories to Understand Student Learning Process with Graph-based Reasonable Knowledge Tracing}. In \bibinfo{booktitle}{\emph{Proceedings of the 30th ACM SIGKDD international conference on knowledge discovery \& data mining}}.
\newblock


\bibitem[Cui et~al\mbox{.}(2024b)]%
        {counterfactual-kt}
\bibfield{author}{\bibinfo{person}{Jiajun Cui}, \bibinfo{person}{Minghe Yu}, \bibinfo{person}{Bo Jiang}, \bibinfo{person}{Aimin Zhou}, \bibinfo{person}{Jianyong Wang}, {and} \bibinfo{person}{Wei Zhang}.} \bibinfo{year}{2024}\natexlab{b}.
\newblock \showarticletitle{Interpretable Knowledge Tracing via Response Influence-based Counterfactual Reasoning}. In \bibinfo{booktitle}{\emph{Proceedings of the 40th IEEE International Conference on Data Engineering}}.
\newblock


\bibitem[Cui et~al\mbox{.}(2019)]%
        {class-balance}
\bibfield{author}{\bibinfo{person}{Yin Cui}, \bibinfo{person}{Menglin Jia}, \bibinfo{person}{Tsung-Yi Lin}, \bibinfo{person}{Yang Song}, {and} \bibinfo{person}{Serge Belongie}.} \bibinfo{year}{2019}\natexlab{}.
\newblock \showarticletitle{Class-balanced loss based on effective number of samples}. In \bibinfo{booktitle}{\emph{Proceedings of the IEEE/CVF conference on computer vision and pattern recognition}}. \bibinfo{pages}{9268--9277}.
\newblock


\bibitem[Douzas et~al\mbox{.}(2018)]%
        {douzas_smote}
\bibfield{author}{\bibinfo{person}{Georgios Douzas}, \bibinfo{person}{Fernando Bacao}, {and} \bibinfo{person}{Felix Last}.} \bibinfo{year}{2018}\natexlab{}.
\newblock \showarticletitle{Improving imbalanced learning through a heuristic oversampling method based on k-means and SMOTE}.
\newblock \bibinfo{journal}{\emph{Information Sciences}}  \bibinfo{volume}{465} (\bibinfo{year}{2018}), \bibinfo{pages}{1--20}.
\newblock


\bibitem[Embretson and Reise(2013)]%
        {irt}
\bibfield{author}{\bibinfo{person}{Susan~E Embretson} {and} \bibinfo{person}{Steven~P Reise}.} \bibinfo{year}{2013}\natexlab{}.
\newblock \bibinfo{booktitle}{\emph{Item response theory}}.
\newblock \bibinfo{publisher}{Psychology Press}.
\newblock


\bibitem[Feng et~al\mbox{.}(2009)]%
        {assistments}
\bibfield{author}{\bibinfo{person}{Mingyu Feng}, \bibinfo{person}{Neil Heffernan}, {and} \bibinfo{person}{Kenneth Koedinger}.} \bibinfo{year}{2009}\natexlab{}.
\newblock \showarticletitle{Addressing the assessment challenge with an online system that tutors as it assesses}.
\newblock \bibinfo{journal}{\emph{User modeling and user-adapted interaction}}  \bibinfo{volume}{19} (\bibinfo{year}{2009}), \bibinfo{pages}{243--266}.
\newblock


\bibitem[Ghosh et~al\mbox{.}(2020)]%
        {akt}
\bibfield{author}{\bibinfo{person}{Aritra Ghosh}, \bibinfo{person}{Neil Heffernan}, {and} \bibinfo{person}{Andrew~S Lan}.} \bibinfo{year}{2020}\natexlab{}.
\newblock \showarticletitle{Context-aware attentive knowledge tracing}. In \bibinfo{booktitle}{\emph{Proceedings of the 26th ACM SIGKDD international conference on knowledge discovery \& data mining}}. \bibinfo{pages}{2330--2339}.
\newblock


\bibitem[Gupta et~al\mbox{.}(2022)]%
        {gupta2022cse}
\bibfield{author}{\bibinfo{person}{Neha Gupta}, \bibinfo{person}{Vinita Jindal}, {and} \bibinfo{person}{Punam Bedi}.} \bibinfo{year}{2022}\natexlab{}.
\newblock \showarticletitle{CSE-IDS: Using cost-sensitive deep learning and ensemble algorithms to handle class imbalance in network-based intrusion detection systems}.
\newblock \bibinfo{journal}{\emph{Computers \& Security}}  \bibinfo{volume}{112} (\bibinfo{year}{2022}), \bibinfo{pages}{102499}.
\newblock


\bibitem[Haixiang et~al\mbox{.}(2017)]%
        {imbalance_review}
\bibfield{author}{\bibinfo{person}{Guo Haixiang}, \bibinfo{person}{Li Yijing}, \bibinfo{person}{Jennifer Shang}, \bibinfo{person}{Gu Mingyun}, \bibinfo{person}{Huang Yuanyue}, {and} \bibinfo{person}{Gong Bing}.} \bibinfo{year}{2017}\natexlab{}.
\newblock \showarticletitle{Learning from class-imbalanced data: Review of methods and applications}.
\newblock \bibinfo{journal}{\emph{Expert systems with applications}}  \bibinfo{volume}{73} (\bibinfo{year}{2017}), \bibinfo{pages}{220--239}.
\newblock


\bibitem[Han et~al\mbox{.}(2005)]%
        {han_smote}
\bibfield{author}{\bibinfo{person}{Hui Han}, \bibinfo{person}{Wen-Yuan Wang}, {and} \bibinfo{person}{Bing-Huan Mao}.} \bibinfo{year}{2005}\natexlab{}.
\newblock \showarticletitle{Borderline-SMOTE: a new over-sampling method in imbalanced data sets learning}. In \bibinfo{booktitle}{\emph{International conference on intelligent computing}}. Springer, \bibinfo{pages}{878--887}.
\newblock


\bibitem[Huang et~al\mbox{.}(2020)]%
        {huang2020learning}
\bibfield{author}{\bibinfo{person}{Zhenya Huang}, \bibinfo{person}{Qi Liu}, \bibinfo{person}{Yuying Chen}, \bibinfo{person}{Le Wu}, \bibinfo{person}{Keli Xiao}, \bibinfo{person}{Enhong Chen}, \bibinfo{person}{Haiping Ma}, {and} \bibinfo{person}{Guoping Hu}.} \bibinfo{year}{2020}\natexlab{}.
\newblock \showarticletitle{Learning or forgetting? a dynamic approach for tracking the knowledge proficiency of students}.
\newblock \bibinfo{journal}{\emph{ACM Transactions on Information Systems (TOIS)}} \bibinfo{volume}{38}, \bibinfo{number}{2} (\bibinfo{year}{2020}), \bibinfo{pages}{1--33}.
\newblock


\bibitem[K{\"a}ser et~al\mbox{.}(2017)]%
        {dbn}
\bibfield{author}{\bibinfo{person}{Tanja K{\"a}ser}, \bibinfo{person}{Severin Klingler}, \bibinfo{person}{Alexander~G Schwing}, {and} \bibinfo{person}{Markus Gross}.} \bibinfo{year}{2017}\natexlab{}.
\newblock \showarticletitle{Dynamic Bayesian networks for student modeling}.
\newblock \bibinfo{journal}{\emph{IEEE Transactions on Learning Technologies}} \bibinfo{volume}{10}, \bibinfo{number}{4} (\bibinfo{year}{2017}), \bibinfo{pages}{450--462}.
\newblock


\bibitem[Kingma and Ba(2014)]%
        {adam}
\bibfield{author}{\bibinfo{person}{Diederik~P Kingma} {and} \bibinfo{person}{Jimmy Ba}.} \bibinfo{year}{2014}\natexlab{}.
\newblock \showarticletitle{Adam: A method for stochastic optimization}.
\newblock \bibinfo{journal}{\emph{arXiv preprint arXiv:1412.6980}} (\bibinfo{year}{2014}).
\newblock


\bibitem[Krawczyk et~al\mbox{.}(2014)]%
        {cost_tree}
\bibfield{author}{\bibinfo{person}{Bartosz Krawczyk}, \bibinfo{person}{Micha{\l} Wo{\'z}niak}, {and} \bibinfo{person}{Gerald Schaefer}.} \bibinfo{year}{2014}\natexlab{}.
\newblock \showarticletitle{Cost-sensitive decision tree ensembles for effective imbalanced classification}.
\newblock \bibinfo{journal}{\emph{Applied Soft Computing}}  \bibinfo{volume}{14} (\bibinfo{year}{2014}), \bibinfo{pages}{554--562}.
\newblock


\bibitem[LeCun et~al\mbox{.}(2015)]%
        {deep-learning}
\bibfield{author}{\bibinfo{person}{Yann LeCun}, \bibinfo{person}{Yoshua Bengio}, {and} \bibinfo{person}{Geoffrey Hinton}.} \bibinfo{year}{2015}\natexlab{}.
\newblock \showarticletitle{Deep learning}.
\newblock \bibinfo{journal}{\emph{nature}} \bibinfo{volume}{521}, \bibinfo{number}{7553} (\bibinfo{year}{2015}), \bibinfo{pages}{436--444}.
\newblock


\bibitem[Lee et~al\mbox{.}(2022)]%
        {clkt}
\bibfield{author}{\bibinfo{person}{Wonsung Lee}, \bibinfo{person}{Jaeyoon Chun}, \bibinfo{person}{Youngmin Lee}, \bibinfo{person}{Kyoungsoo Park}, {and} \bibinfo{person}{Sungrae Park}.} \bibinfo{year}{2022}\natexlab{}.
\newblock \showarticletitle{Contrastive learning for knowledge tracing}. In \bibinfo{booktitle}{\emph{Proceedings of the ACM Web Conference 2022}}. \bibinfo{pages}{2330--2338}.
\newblock


\bibitem[Li et~al\mbox{.}(2019)]%
        {ghm}
\bibfield{author}{\bibinfo{person}{Buyu Li}, \bibinfo{person}{Yu Liu}, {and} \bibinfo{person}{Xiaogang Wang}.} \bibinfo{year}{2019}\natexlab{}.
\newblock \showarticletitle{Gradient harmonized single-stage detector}. In \bibinfo{booktitle}{\emph{Proceedings of the AAAI conference on artificial intelligence}}, Vol.~\bibinfo{volume}{33}. \bibinfo{pages}{8577--8584}.
\newblock


\bibitem[Li et~al\mbox{.}(2020)]%
        {dice}
\bibfield{author}{\bibinfo{person}{Xiaoya Li}, \bibinfo{person}{Xiaofei Sun}, \bibinfo{person}{Yuxian Meng}, \bibinfo{person}{Junjun Liang}, \bibinfo{person}{Fei Wu}, {and} \bibinfo{person}{Jiwei Li}.} \bibinfo{year}{2020}\natexlab{}.
\newblock \showarticletitle{Dice Loss for Data-imbalanced NLP Tasks}. In \bibinfo{booktitle}{\emph{Proceedings of the 58th Annual Meeting of the Association for Computational Linguistics}}. \bibinfo{pages}{465--476}.
\newblock


\bibitem[Lin et~al\mbox{.}(2017)]%
        {focal}
\bibfield{author}{\bibinfo{person}{Tsung-Yi Lin}, \bibinfo{person}{Priya Goyal}, \bibinfo{person}{Ross Girshick}, \bibinfo{person}{Kaiming He}, {and} \bibinfo{person}{Piotr Doll{\'a}r}.} \bibinfo{year}{2017}\natexlab{}.
\newblock \showarticletitle{Focal loss for dense object detection}. In \bibinfo{booktitle}{\emph{Proceedings of the IEEE international conference on computer vision}}. \bibinfo{pages}{2980--2988}.
\newblock


\bibitem[Liu et~al\mbox{.}(2019)]%
        {ekt}
\bibfield{author}{\bibinfo{person}{Qi Liu}, \bibinfo{person}{Zhenya Huang}, \bibinfo{person}{Yu Yin}, \bibinfo{person}{Enhong Chen}, \bibinfo{person}{Hui Xiong}, \bibinfo{person}{Yu Su}, {and} \bibinfo{person}{Guoping Hu}.} \bibinfo{year}{2019}\natexlab{}.
\newblock \showarticletitle{Ekt: Exercise-aware knowledge tracing for student performance prediction}.
\newblock \bibinfo{journal}{\emph{IEEE Transactions on Knowledge and Data Engineering}} \bibinfo{volume}{33}, \bibinfo{number}{1} (\bibinfo{year}{2019}), \bibinfo{pages}{100--115}.
\newblock


\bibitem[Liu et~al\mbox{.}(2008)]%
        {undersampling}
\bibfield{author}{\bibinfo{person}{Xu-Ying Liu}, \bibinfo{person}{Jianxin Wu}, {and} \bibinfo{person}{Zhi-Hua Zhou}.} \bibinfo{year}{2008}\natexlab{}.
\newblock \showarticletitle{Exploratory undersampling for class-imbalance learning}.
\newblock \bibinfo{journal}{\emph{IEEE Transactions on Systems, Man, and Cybernetics, Part B (Cybernetics)}} \bibinfo{volume}{39}, \bibinfo{number}{2} (\bibinfo{year}{2008}), \bibinfo{pages}{539--550}.
\newblock


\bibitem[Liu et~al\mbox{.}(2023)]%
        {Liu2023}
\bibfield{author}{\bibinfo{person}{Yingjie Liu}, \bibinfo{person}{Tiancheng Zhang}, \bibinfo{person}{Xuecen Wang}, \bibinfo{person}{Ge Yu}, {and} \bibinfo{person}{Tao Li}.} \bibinfo{year}{2023}\natexlab{}.
\newblock \showarticletitle{New development of cognitive diagnosis models}.
\newblock \bibinfo{journal}{\emph{Frontiers of Computer Science}} \bibinfo{volume}{17}, \bibinfo{number}{1} (\bibinfo{year}{2023}), \bibinfo{pages}{171604}.
\newblock


\bibitem[Long et~al\mbox{.}(2021)]%
        {iekt}
\bibfield{author}{\bibinfo{person}{Ting Long}, \bibinfo{person}{Yunfei Liu}, \bibinfo{person}{Jian Shen}, \bibinfo{person}{Weinan Zhang}, {and} \bibinfo{person}{Yong Yu}.} \bibinfo{year}{2021}\natexlab{}.
\newblock \showarticletitle{Tracing knowledge state with individual cognition and acquisition estimation}. In \bibinfo{booktitle}{\emph{Proceedings of the 44th International ACM SIGIR Conference on Research and Development in Information Retrieval}}. \bibinfo{pages}{173--182}.
\newblock


\bibitem[L{\'o}pez et~al\mbox{.}(2015)]%
        {lopez2015cost}
\bibfield{author}{\bibinfo{person}{Victoria L{\'o}pez}, \bibinfo{person}{Sara Del~R{\'\i}o}, \bibinfo{person}{Jos{\'e}~Manuel Ben{\'\i}tez}, {and} \bibinfo{person}{Francisco Herrera}.} \bibinfo{year}{2015}\natexlab{}.
\newblock \showarticletitle{Cost-sensitive linguistic fuzzy rule based classification systems under the MapReduce framework for imbalanced big data}.
\newblock \bibinfo{journal}{\emph{Fuzzy Sets and Systems}}  \bibinfo{volume}{258} (\bibinfo{year}{2015}), \bibinfo{pages}{5--38}.
\newblock


\bibitem[Maldonado and L{\'o}pez(2014)]%
        {maldonado2014imbalanced}
\bibfield{author}{\bibinfo{person}{Sebasti{\'a}n Maldonado} {and} \bibinfo{person}{Julio L{\'o}pez}.} \bibinfo{year}{2014}\natexlab{}.
\newblock \showarticletitle{Imbalanced data classification using second-order cone programming support vector machines}.
\newblock \bibinfo{journal}{\emph{Pattern Recognition}} \bibinfo{volume}{47}, \bibinfo{number}{5} (\bibinfo{year}{2014}), \bibinfo{pages}{2070--2079}.
\newblock


\bibitem[Milletari et~al\mbox{.}(2016)]%
        {vnet}
\bibfield{author}{\bibinfo{person}{Fausto Milletari}, \bibinfo{person}{Nassir Navab}, {and} \bibinfo{person}{Seyed-Ahmad Ahmadi}.} \bibinfo{year}{2016}\natexlab{}.
\newblock \showarticletitle{V-net: Fully convolutional neural networks for volumetric medical image segmentation}. In \bibinfo{booktitle}{\emph{2016 fourth international conference on 3D vision (3DV)}}. \bibinfo{pages}{565--571}.
\newblock


\bibitem[Nagatani et~al\mbox{.}(2019)]%
        {dkt-forget}
\bibfield{author}{\bibinfo{person}{Koki Nagatani}, \bibinfo{person}{Qian Zhang}, \bibinfo{person}{Masahiro Sato}, \bibinfo{person}{Yan-Ying Chen}, \bibinfo{person}{Francine Chen}, {and} \bibinfo{person}{Tomoko Ohkuma}.} \bibinfo{year}{2019}\natexlab{}.
\newblock \showarticletitle{Augmenting knowledge tracing by considering forgetting behavior}. In \bibinfo{booktitle}{\emph{The world wide web conference}}. \bibinfo{pages}{3101--3107}.
\newblock


\bibitem[Pad{\'o}(2017)]%
        {norming}
\bibfield{author}{\bibinfo{person}{Ulrike Pad{\'o}}.} \bibinfo{year}{2017}\natexlab{}.
\newblock \showarticletitle{Question difficulty--how to estimate without norming, how to use for automated grading}. In \bibinfo{booktitle}{\emph{Proceedings of the 12th workshop on innovative use of NLP for building educational applications}}. \bibinfo{pages}{1--10}.
\newblock


\bibitem[Pandey and Karypis(2019)]%
        {sakt}
\bibfield{author}{\bibinfo{person}{Shalini Pandey} {and} \bibinfo{person}{George Karypis}.} \bibinfo{year}{2019}\natexlab{}.
\newblock \showarticletitle{A self-attentive model for knowledge tracing}. In \bibinfo{booktitle}{\emph{EDM 2019 - Proceedings of the 12th International Conference on Educational Data Mining}}. \bibinfo{pages}{384--389}.
\newblock


\bibitem[Pandey and Srivastava(2020)]%
        {rkt}
\bibfield{author}{\bibinfo{person}{Shalini Pandey} {and} \bibinfo{person}{Jaideep Srivastava}.} \bibinfo{year}{2020}\natexlab{}.
\newblock \showarticletitle{RKT: relation-aware self-attention for knowledge tracing}. In \bibinfo{booktitle}{\emph{Proceedings of the 29th ACM International Conference on Information \& Knowledge Management}}. \bibinfo{pages}{1205--1214}.
\newblock


\bibitem[Pardos and Heffernan(2010)]%
        {ibkt}
\bibfield{author}{\bibinfo{person}{Zachary~A Pardos} {and} \bibinfo{person}{Neil~T Heffernan}.} \bibinfo{year}{2010}\natexlab{}.
\newblock \showarticletitle{Modeling individualization in a bayesian networks implementation of knowledge tracing}. In \bibinfo{booktitle}{\emph{User Modeling, Adaptation, and Personalization: 18th International Conference, UMAP 2010, Big Island, HI, USA, June 20-24, 2010. Proceedings 18}}. \bibinfo{pages}{255--266}.
\newblock


\bibitem[Pardos and Heffernan(2011)]%
        {kt-idem}
\bibfield{author}{\bibinfo{person}{Zachary~A Pardos} {and} \bibinfo{person}{Neil~T Heffernan}.} \bibinfo{year}{2011}\natexlab{}.
\newblock \showarticletitle{KT-IDEM: Introducing item difficulty to the knowledge tracing model}. In \bibinfo{booktitle}{\emph{User Modeling, Adaption and Personalization: 19th International Conference, UMAP 2011, Girona, Spain, July 11-15, 2011. Proceedings 19}}. \bibinfo{pages}{243--254}.
\newblock


\bibitem[Piech et~al\mbox{.}(2015)]%
        {dkt}
\bibfield{author}{\bibinfo{person}{Chris Piech}, \bibinfo{person}{Jonathan Bassen}, \bibinfo{person}{Jonathan Huang}, \bibinfo{person}{Surya Ganguli}, \bibinfo{person}{Mehran Sahami}, \bibinfo{person}{Leonidas~J Guibas}, {and} \bibinfo{person}{Jascha Sohl-Dickstein}.} \bibinfo{year}{2015}\natexlab{}.
\newblock \showarticletitle{Deep Knowledge Tracing}. In \bibinfo{booktitle}{\emph{Advances in Neural Information Processing Systems}}, Vol.~\bibinfo{volume}{28}.
\newblock


\bibitem[Rosenbaum(1987)]%
        {ipw}
\bibfield{author}{\bibinfo{person}{Paul~R Rosenbaum}.} \bibinfo{year}{1987}\natexlab{}.
\newblock \showarticletitle{Model-based direct adjustment}.
\newblock \bibinfo{journal}{\emph{Journal of the American statistical Association}} \bibinfo{volume}{82}, \bibinfo{number}{398} (\bibinfo{year}{1987}), \bibinfo{pages}{387--394}.
\newblock


\bibitem[Schnabel et~al\mbox{.}(2016)]%
        {rec-debias}
\bibfield{author}{\bibinfo{person}{Tobias Schnabel}, \bibinfo{person}{Adith Swaminathan}, \bibinfo{person}{Ashudeep Singh}, \bibinfo{person}{Navin Chandak}, {and} \bibinfo{person}{Thorsten Joachims}.} \bibinfo{year}{2016}\natexlab{}.
\newblock \showarticletitle{Recommendations as treatments: Debiasing learning and evaluation}. In \bibinfo{booktitle}{\emph{international conference on machine learning}}. \bibinfo{pages}{1670--1679}.
\newblock


\bibitem[Shen et~al\mbox{.}(2024)]%
        {shen2024}
\bibfield{author}{\bibinfo{person}{Junhao Shen}, \bibinfo{person}{Hong Qian}, \bibinfo{person}{Wei Zhang}, {and} \bibinfo{person}{Aimin Zhou}.} \bibinfo{year}{2024}\natexlab{}.
\newblock \showarticletitle{Symbolic Cognitive Diagnosis via Hybrid Optimization for Intelligent Education Systems}. In \bibinfo{booktitle}{\emph{Proceedings of the Thirty-Eighth {AAAI} Conference on Artificial Intelligence}}. \bibinfo{address}{Vancouver, Canada}, \bibinfo{pages}{14928--14936}.
\newblock


\bibitem[Shen et~al\mbox{.}(2022)]%
        {dimkt}
\bibfield{author}{\bibinfo{person}{Shuanghong Shen}, \bibinfo{person}{Zhenya Huang}, \bibinfo{person}{Qi Liu}, \bibinfo{person}{Yu Su}, \bibinfo{person}{Shijin Wang}, {and} \bibinfo{person}{Enhong Chen}.} \bibinfo{year}{2022}\natexlab{}.
\newblock \showarticletitle{Assessing Student's Dynamic Knowledge State by Exploring the Question Difficulty Effect}. In \bibinfo{booktitle}{\emph{Proceedings of the 45th International ACM SIGIR Conference on Research and Development in Information Retrieval}}. \bibinfo{pages}{427--437}.
\newblock


\bibitem[Shen et~al\mbox{.}(2021)]%
        {lpkt}
\bibfield{author}{\bibinfo{person}{Shuanghong Shen}, \bibinfo{person}{Qi Liu}, \bibinfo{person}{Enhong Chen}, \bibinfo{person}{Zhenya Huang}, \bibinfo{person}{Wei Huang}, \bibinfo{person}{Yu Yin}, \bibinfo{person}{Yu Su}, {and} \bibinfo{person}{Shijin Wang}.} \bibinfo{year}{2021}\natexlab{}.
\newblock \showarticletitle{Learning process-consistent knowledge tracing}. In \bibinfo{booktitle}{\emph{Proceedings of the 27th ACM SIGKDD conference on knowledge discovery \& data mining}}. \bibinfo{pages}{1452--1460}.
\newblock


\bibitem[Sim and Rasiah(2006)]%
        {item-diff-disc}
\bibfield{author}{\bibinfo{person}{Si-Mui Sim} {and} \bibinfo{person}{Raja~Isaiah Rasiah}.} \bibinfo{year}{2006}\natexlab{}.
\newblock \showarticletitle{Relationship between item difficulty and discrimination indices in true/false-type multiple choice questions of a para-clinical multidisciplinary paper}.
\newblock \bibinfo{journal}{\emph{Annals-Academy of Medicine Singapore}} \bibinfo{volume}{35}, \bibinfo{number}{2} (\bibinfo{year}{2006}), \bibinfo{pages}{67}.
\newblock


\bibitem[Solomatine and Ostfeld(2008)]%
        {data-driven}
\bibfield{author}{\bibinfo{person}{Dimitri~P Solomatine} {and} \bibinfo{person}{Avi Ostfeld}.} \bibinfo{year}{2008}\natexlab{}.
\newblock \showarticletitle{Data-driven modelling: some past experiences and new approaches}.
\newblock \bibinfo{journal}{\emph{Journal of hydroinformatics}} \bibinfo{volume}{10}, \bibinfo{number}{1} (\bibinfo{year}{2008}), \bibinfo{pages}{3--22}.
\newblock


\bibitem[Sun et~al\mbox{.}(2015)]%
        {sun_undersampling}
\bibfield{author}{\bibinfo{person}{Zhongbin Sun}, \bibinfo{person}{Qinbao Song}, \bibinfo{person}{Xiaoyan Zhu}, \bibinfo{person}{Heli Sun}, \bibinfo{person}{Baowen Xu}, {and} \bibinfo{person}{Yuming Zhou}.} \bibinfo{year}{2015}\natexlab{}.
\newblock \showarticletitle{A novel ensemble method for classifying imbalanced data}.
\newblock \bibinfo{journal}{\emph{Pattern Recognition}} \bibinfo{volume}{48}, \bibinfo{number}{5} (\bibinfo{year}{2015}), \bibinfo{pages}{1623--1637}.
\newblock


\bibitem[Vaswani et~al\mbox{.}(2017)]%
        {attention}
\bibfield{author}{\bibinfo{person}{Ashish Vaswani}, \bibinfo{person}{Noam Shazeer}, \bibinfo{person}{Niki Parmar}, \bibinfo{person}{Jakob Uszkoreit}, \bibinfo{person}{Llion Jones}, \bibinfo{person}{Aidan~N Gomez}, \bibinfo{person}{{\L}ukasz Kaiser}, {and} \bibinfo{person}{Illia Polosukhin}.} \bibinfo{year}{2017}\natexlab{}.
\newblock \showarticletitle{Attention is all you need}.
\newblock \bibinfo{journal}{\emph{Advances in neural information processing systems}}  \bibinfo{volume}{30} (\bibinfo{year}{2017}).
\newblock


\bibitem[Wang et~al\mbox{.}(2021)]%
        {hawkesKT}
\bibfield{author}{\bibinfo{person}{Chenyang Wang}, \bibinfo{person}{Weizhi Ma}, \bibinfo{person}{Min Zhang}, \bibinfo{person}{Chuancheng Lv}, \bibinfo{person}{Fengyuan Wan}, \bibinfo{person}{Huijie Lin}, \bibinfo{person}{Taoran Tang}, \bibinfo{person}{Yiqun Liu}, {and} \bibinfo{person}{Shaoping Ma}.} \bibinfo{year}{2021}\natexlab{}.
\newblock \showarticletitle{Temporal cross-effects in knowledge tracing}. In \bibinfo{booktitle}{\emph{Proceedings of the 14th ACM International Conference on Web Search and Data Mining}}. \bibinfo{pages}{517--525}.
\newblock


\bibitem[Wang et~al\mbox{.}(2018)]%
        {position}
\bibfield{author}{\bibinfo{person}{Xuanhui Wang}, \bibinfo{person}{Nadav Golbandi}, \bibinfo{person}{Michael Bendersky}, \bibinfo{person}{Donald Metzler}, {and} \bibinfo{person}{Marc Najork}.} \bibinfo{year}{2018}\natexlab{}.
\newblock \showarticletitle{Position bias estimation for unbiased learning to rank in personal search}. In \bibinfo{booktitle}{\emph{Proceedings of the eleventh ACM international conference on web search and data mining}}. \bibinfo{pages}{610--618}.
\newblock


\bibitem[Wang et~al\mbox{.}(2020)]%
        {eedi}
\bibfield{author}{\bibinfo{person}{Zichao Wang}, \bibinfo{person}{Angus Lamb}, \bibinfo{person}{Evgeny Saveliev}, \bibinfo{person}{Pashmina Cameron}, \bibinfo{person}{Yordan Zaykov}, \bibinfo{person}{Jos{\'e}~Miguel Hern{\'a}ndez-Lobato}, \bibinfo{person}{Richard~E Turner}, \bibinfo{person}{Richard~G Baraniuk}, \bibinfo{person}{Craig Barton}, \bibinfo{person}{Simon~Peyton Jones}, {et~al\mbox{.}}} \bibinfo{year}{2020}\natexlab{}.
\newblock \showarticletitle{Instructions and guide for diagnostic questions: The neurips 2020 education challenge}.
\newblock \bibinfo{journal}{\emph{arXiv preprint arXiv:2007.12061}} (\bibinfo{year}{2020}).
\newblock


\bibitem[Zhang et~al\mbox{.}(2016)]%
        {zhang_undersampling}
\bibfield{author}{\bibinfo{person}{Zhongliang Zhang}, \bibinfo{person}{Bartosz Krawczyk}, \bibinfo{person}{Salvador Garcia}, \bibinfo{person}{Alejandro Rosales-P{\'e}rez}, {and} \bibinfo{person}{Francisco Herrera}.} \bibinfo{year}{2016}\natexlab{}.
\newblock \showarticletitle{Empowering one-vs-one decomposition with ensemble learning for multi-class imbalanced data}.
\newblock \bibinfo{journal}{\emph{Knowledge-Based Systems}}  \bibinfo{volume}{106} (\bibinfo{year}{2016}), \bibinfo{pages}{251--263}.
\newblock


\end{thebibliography}


\end{document}